\documentclass[aps,reprint,amssymb,onecolumn,nofootinbib]{revtex4-1}
\pdfoutput=1

\usepackage{natbib}
\usepackage{grffile}

\usepackage{dsfont}
\usepackage{amsmath}
\usepackage{graphicx}
\usepackage[english]{babel}

\usepackage{color}
\usepackage{bm}
\usepackage{comment}
 \usepackage{dutchcal}
\usepackage{mathrsfs}
\usepackage{booktabs}

\newcommand{\modif}[1]{{\bf #1}}
\renewcommand{\modif}[1]{{ #1}}
\begin{document}

\newcommand{\bA}{{\bf A}}
\newcommand{\bX}{{\bf X}}
\renewcommand {\paragraph}[1]{{\bf #1 }}

\newcommand {\av}[1]{\left\langle #1 \right \rangle}
\newcommand{\bx}{{\bf x}}
\newcommand{\bk}{{\bf k}}
\newcommand{\br}{{\bf r}}
\newcommand{\bv}{{\bf v}}
\newcommand{\bz}{{\bf z}}

\newcommand{\bbeta}{{\boldsymbol \beta}}
\newcommand{\bmW}{{\boldsymbol {\mathcal W}}}
\newcommand{\blambda}{{\boldsymbol \lambda}}
\newcommand{\bchi}{{\boldsymbol \chi}}
\newcommand{\brho}{{\boldsymbol \rho}}
\newcommand{\kmax}{{k_{\text{max}}}}
\newcommand{\dsp}{\displaystyle}

\newcommand{\mC}{\mathcal C}
\newcommand{\mD}{\mathcal D}
\newcommand{\mH}{\mathcal H}
\newcommand{\mJ}{\mathcal J}
\newcommand{\mL}{\mathcal L}
\newcommand{\mM}{\mathcal M}
\newcommand{\mO}{\mathcal O}
\newcommand{\mV}{\mathcal V}
\newcommand{\mW}{\mathcal W}
\newcommand{\mX}{\mathcal X}
\newcommand{\mZ}{\mathcal Z}

\newcommand{\mbR}{\mathbf R}

\newcommand{\ps}{p_{\text{stat}}}
\newcommand{\ds}{\text{ds}}
\newcommand{\dt}{\text{dt}}
\newcommand{\dx}{\text{dx}}
\newcommand{\dy}{\text{\bf dy}}
\newcommand{\dd}{\text{d}}

\newcommand{\zj}[2]{j_{#1}^{(#2)}}
%\zj(m,n) = n ieme zero de la fonction de bessel de premiere espece J d'ordre m

\newcommand{\dbx}{\text{\bf dx}}

\newcommand{\rhs}{\text{rhs}}
\newcommand{\wit}{WIT}
\newcommand{\cit}{CIT}
\newcommand{\cp}{\wp }

\newcommand{\ball}{\text{bal}}
\newcommand{\puff}{\text{B/E}}

\newcommand{\mc}{\mathcal c}

\newcommand{\bla}[1]{\textcolor{blue}{#1}}
\title{Turbulent mass inhomogeneities induced by a point-source }
\author{Simon Thalabard}
\email{simon.thalabard@ens-lyon.org}
\affiliation{Laboratoire Lagrange, Universit\'e C\^ote d'Azur,\\ CNRS, Observatoire de la C\^ote d'Azur,  Bd.\ de l'Observatoire, 06300 Nice, France.}
\date{\today}
\begin{abstract}
We describe how turbulence distributes tracers away from a localized source of injection, and analyze how the spatial inhomogeneities of the concentration field depend on the amount of randomness in the injection mechanism. For that purpose, we contrast the mass correlations induced by  purely random injections  with those induced by  continuous injections in the environment. Using the Kraichnan model of turbulent advection, whereby the underlying velocity field is assumed  to be shortly correlated in time, we explicitly identify scaling regions for the statistics of the mass contained within a shell of radius $r$ and located at a distance $\rho$ away from the source. The two key parameters are found to be \emph{(i)} the ratio $s^2$ between  the absolute and the relative  timescales of dispersion and \emph{(ii)} the ratio $\Lambda$ between the size of the cloud and its distance away from the source. When the injection is random, only the former is relevant, as previously shown by Celani, Martins-Afonso \& Mazzino, \emph{J. Fluid Mech}, 2007 in the case of an incompressible fluid. It is argued that the space partition in terms of  $s^2$ and $\Lambda$ is a robust feature of the injection mechanism itself,  which should remain relevant beyond the Kraichnan model. 
This is for instance the case in a generalized version of the  model,  where the absolute dispersion is prescribed to be ballistic rather than diffusive. 
\end{abstract}

\maketitle

\section{Introduction}
Modeling  how tracer particles released from a source  distribute in a turbulent environment  is a long-standing problem with obvious practical implications, from hazard control to biology \cite{csanady_turbulent_2012,nasstrom_national_2007,devenish_sensitivity_2012,celani_odor_2014}. In such a scenario, the statistics of the concentration field are non-homogeneous, and obviously depend on  the space-time statistics of both  the  underlying turbulent velocity field and of the injection mechanism. In   the case of a nearly-punctual source, the latter reduces  to the specific features  of the time statistics of the injection.
In the engineering community, those matters have led to the development and systematic use of Lagrangian stochastic models, whose versatility to accommodate both various turbulent statistics and various types of source make them appropriate for commercial use \cite{flesch_backward-time_1995,wilson_review_1996,jones_uk_2004}. 
From a more fundamental perspective, quantitative estimates are however seldom, that study how the spatial distribution of the  concentration depends upon the properties of the flow field, and its interplay with the injection statistics. 
Among exceptions are the papers \cite{celani_point-source_2007}  and \cite{afonso_point-source_2011}, where the advecting velocity field is assumed to have a vanishing correlation time  and is  prescribed by the ``Kraichnan  ensemble''  (later defined). More specifically,  the authors in \cite{celani_point-source_2007} consider a memoryless point-source that emits randomly in time, and show that the two-point correlation of the concentration field exhibit  non-trivial intermittent scaling regimes, that are essentially determined by the ratio between the absolute and relative dispersion timescales.

In practice, this kind of a random ``source''  only  \emph{modulates} the number of particles, but does not on average inject any matter in the environment. To model a genuine source, one therefore needs to superimpose over the modulation a continuous contribution, that determines the net injection rate. The presence of a continuous contribution might however fundamentally alter the spatial patterns of the concentration field away from the source, which are determined by an intricate interplay between the turbulent environment  and the injection statistics. The Lagrangian point of view, ties the statistics of order $n$ of the concentration field to Lagrangian averages over $n$ distinct trajectories (see \emph{e.g.} \cite{celani_statistical_2001}) -- The statistical ensemble being determined by the injection mechanism itself. 

\modif{For the white-in-time point-source considered in  \cite{celani_point-source_2007}~: the $n$-point spatial  correlations are in principle determined by the Lagrangian statistics for  puffs of ($k\le n$) tracers that transit  simultaneously through the source \cite{celani_odor_2014}.  While the statistics of such turbulent puffs may prove intricate \cite{bianchi_evolution_2016}, the spatial two-point correlations patterns induced by white-in-time injections remain simple in the sense that they relate to standard two-time Lagrangian statistics.

On the other hand, when the injection is continuous, every Lagrangian trajectory transiting through the source (regardless when) contributes to the statistics. In the point-source setting the Lagrangian statistics involved to determine the $n$-point statistics are therefore non-standard $ N+1$ Lagrangian statistics --  involving the $N$ different times of injections and the measurement time. In this heuristic picture, a continuously emitting point-source therefore mixes many Lagrangian time-scales , and it is not clear whether scaling should be expected at all when it comes to the statistics  of the concentration field (see for instance \cite[Chapter 3]{laenen_modulation_2017}).}

One purpose of the present paper is to clarify this issue, and to contrast in a quantitative fashion how the concentration statistics depend on  the injection mechanism, and in particular on the amount of randomness in the injection statistics. To do so, we contrast the spatial correlations of the concentration fields induced by a  
localized white-in-time modulation (later \wit) to those induced by a localized continuous-in-time injection (later \cit). The concentration depends linearly upon the source statistics, and  the full correlation induced by a source with both continuous and white components  is recovered  by adding together the   \wit\, and the  \cit\, contributions.
Another purpose of the paper is to highlight the non-trivial dynamical interplay between  absolute and relative dispersion in determining the statistics of the concentration field.

The present work builds on the paper  by \cite{celani_point-source_2007}, but discusses two types of prototypical velocity fields. The first is naturally the Kraichnan ensemble, whose dynamical interplay between  a diffusive absolute and a super-diffusive (Richardson-like) relative dispersion  yet turns out to be highly unrepresentative of a genuine turbulent flow. The second is obtained by slightly altering the Kraichnan velocity field in the spirit of  the so-called ``puff-particles models'' of  \cite{haan_puffparticle_1995}, in order to incorporate some large-scale sweeping, as typically found in Navier-Stokes turbulence. In both cases, it is found that the statistics of the mass contained in a cloud of size $r$ located at a distance $\rho$ away from the source are determined by two key parameters, namely \emph{(i)} the ratio $s^2$ between  the absolute and the relative  timescales of dispersion and \emph{(ii)} the ratio $\Lambda$ between the size of the cloud and its distance away from the source. This partition is robust and is independent on the specificity of the velocity field. When they exist, the specific scaling behaviours  are however non-trivial and intrinsically depend upon the velocity statistics.

The paper is organized as follows. The next section gives a qualitative account on the differences between \cit\, and \wit\, injections, and provides some background definitions on the Kraichnan ensemble. The concentration statistics for the Kraichnan ensemble are derived in Section 3. The effects of the large-scale sweeping  are discussed in Section 4. While Section 3 and 4 are rather technical, the reader can refer to  Figure \ref{fig:sketchscaling} and Tables \ref{table:witscaling} \& \ref{table:citscaling} in the conclusion to find  the main analytical results of the paper readily summarized.
\section{Statistics of the concentration}
This section defines  the  averages and the correlations of the concentration field, and provides some insights on the physical picture  behind both random and continuous injections. Some useful  background material  related to Kraichnan velocity ensembles is also recalled.

\subsection{The concentration field.}
Let us  consider a scenario where the  source only operates from an ``initial time'', say $t = 0$, but is before-hands non-active. 
For negative times, the physical domain $\mD = \mbR^d$  contains  a large number of massless particles (``tracers'') that  are advected by a prescribed turbulent velocity field $\bv$ and subject to a small thermal noise. 
Mathematically, this means that the trajectory $\bX^\varpi(t|\bx_0,t_0)$ of a tracer  $\varpi$ that is at $\bx_0$ at  time $t_0<t$ is obtained as a  specific  realisation of the stochastic system~:%(\cite{cardy_non-equilibrium_2008,gawedzki_stochastic_2008})~:
\begin{equation}
%	\begin{split}
	\dd X_i^{\varpi} =  v_i(\bX^{\varpi},t) \dd t+ \sqrt{2 \kappa} \,\, \dd W_i^{\varpi}, %\;\;  \text{where ${\boldsymbol W}$ is a Wiener process.}
%	\end{split}
	\label{eq:stochtraj}
\end{equation}
where ${\boldsymbol W}$ is a Wiener process.
The statistics of $\bv (\cdot,t)$ are prescribed to be homogeneous and isotropic in space, but  need not yet be fully specified.  Note that the subscripts $i \in [1,d]$ relate to spatial coordinates, while the superscript $\varpi$ denotes a realisation of the noise, different for each tracer. 

  At time $t=0$, the particles distribute over the domain according to an equilibrium distribution $n_0$, obtained after averaging the stochastic trajectories (\ref{eq:stochtraj}) over the noise. The number of particles contained within the infinitesimal volume $\dx$ is then  $n_0(\bx) \dx$.
For positive times $t \ge 0$, the source $S$ becomes active and locally injects or removes particles~: the initial equilibrium   concentration field is then altered into 
\begin{equation}
	n(\bx,t) :=  n_0(\bx) + \int_{0}^{t} \dt_0 \int_{\mD} \dbx_0 \,S(\bx_0,t_0)\, p_\kappa(\bx,t|\bx_0,t_0), \\
	\label{eq:instantn}
\end{equation}
where $ p_\kappa(\bx,t|\bx_0,t_0) := \left\langle \delta \left(\bx - \bX^\varpi(t|\bx_0,t_0)\right) \right\rangle_\kappa$ represents the transition probability from $\bx_0$ to $\bx$, when averaging the tracer trajectories over the noise
\footnote{$\delta(\bx)$ here denotes  the $d$-dimensional Dirac distribution, such that $ \int_{R^d} \dbx \,\delta(\bx)\,\phi(\bx) = \phi(0)$ for any test function $\phi$.}.%
To see that  Formula (\ref{eq:instantn}) indeed defines a quantity that is transported as a density field (\cite{cardy_non-equilibrium_2008}), one needs to use  the invariance property of the equilibrium distribution~: $n_0(\bx) = \int_\mD \dbx_0\, p_\kappa(\bx,t|\bx_0,0) n_0(\bx_0)$, and  observe that the $p_\kappa$'s represent \emph{forward} transition probabilities.\\

\modif{ In this work, we focus on the case of  ``point sources'', whose  spatial extensions $\epsilon$  are taken to be smaller than the integral scale $\lambda$ yet larger than the smallest turbulent scale $\eta_K$ (namely, the dissipation scale in 3D). In order to investigate how the turbulence propagates small-scale inhomogeneities on scales  $\epsilon \ll \delta \ll \lambda$, we will eventually take the three subsequent limits $\eta_K \to 0$,  $\epsilon \to 0$ and then $ \delta \to 0$. 
In order to incorporate some fluctuations in the injection rate, a general source-term could be modeled as ~:}
\begin{equation}
	S_\epsilon(\bx,t) :=  \phi_0 \, \delta_\epsilon(\bx)(1+\sigma \eta(t)),
	\label{eq:realsource}
\end{equation}
\modif{where $\delta_\epsilon$ denotes a compact-support regularization of the Dirac $\delta$  function (see also Paragraph \ref{sec:HankelDetails} for further details). The $\epsilon$ subscripts will later be dropped, and the dependence on the source extension will be made explicit only when necessary.}
$\eta$ is a Gaussian white noise with vanishing mean, that is $\av{\eta}$ = 0  and $\av{\eta(t)\eta(t^\prime)} = \delta(t-t^\prime)$. The injection rate $\phi_0$ is a positive quantity with dimension [time]$^{-1}$. The relative modulation rate $\sigma$ has dimension [time]$^{1/2}$. 

The distinction between \wit\, and \cit\, injections is obtained by isolating in the previous formula the random contribution from the continuous one .   We will therefore study the following simplified emission schemes~:
\begin{equation}
	\begin{split}
	&S(\bx,t) :=  \phi_\sigma \, \delta(\bx) \eta(t) \hspace{0.3cm} \text{(\wit)},\\
	 &\text{or}\;   \hspace{0.2cm}\ S(\bx,t) := \phi_0 \, \delta(\bx) \hspace{0.3cm} \text{(\cit)},
	\label{eq:source}
	\end{split}
\end{equation}
where we write $\phi_\sigma := \phi_0 \sigma$  the  \wit\, modulation rate, with dimension [time]$^{-1/2}$. 
 Let us here again observe, that in this approach, only the \cit\, source term genuinely acts as a source of particles. By contrast, the \wit\, ``source term'' is vanishing on average~: it therefore both acts as a source and a sink of particles,  and only generates \emph{fluctuations} in the distribution in the total number of particles. Naturally, the physical meaning of the \wit\, mechanism is tied to the  the concentration $n(\bx,t)$ not becoming negative.  This constrains the value of  $\phi_\sigma$ to be sufficiently small, so as to guarantee  that the fluctuations remain small with respect to the average equilibrium profile. %We defer for later (Section 3.2) a more quantitative comment on the validity of the \wit\, description.

In the remainder of the paper, we analyse the  steady properties of the concentration field. The \cit\, statistics do not depend on the specific equilibrium distribution $n_0$ and the latter can be safely chosen to be  vanishing.
While this is not so in the \wit\, case, the notation $n(\bx,t)$ will be slightly abused to denote the fluctuation $n(\bx,t)-n_{ref}(\bx)$, with respect to the underlying reference distribution, be the latter the equilibrium or the \cit\, one.

\subsection{Averages and correlations of the concentration field.}
In practice, one wants to describe the mass statistics away from the source, that is the distribution of the mass $\mathcal m(\brho,r,t)$ contained in a ball of diameter $r$ centered at a position $\brho$ (see the sketch in Figure \ref{fig:sketch}):
\begin{equation}
	\mathcal m(\brho,r,t) =  \int_{|\bx - \brho| \le r/2 } \dbx \, n(\bx,t).
\end{equation}
The statistics of the mass is tied to the spatial inhomogeneities of the concentration field: The average mass $\av{ \mathcal m}$ obviously relates to the average concentration $C_1(\bx,t) := \av{n(\bx,t)}$,  while the mass fluctuation $\av{ \mathcal m^2}-  \av{ \mathcal m}^2$ relates to the correlation function  $C_2(\bx,\bx^\prime, t) := \av{n(\bx,t) n(\bx^{\prime},t)}$. Please note, that the averages $\av{\cdot}$ are to be understood in terms of ensemble averages~: over the possible realizations of the turbulent velocity field for the \cit\, case, and over both the turbulent field \emph{and} the source statistics for the \wit\, case. \\
The correlation functions $C_1$ and $C_2$ are the lowest-order non trivial statistics related to the concentration field, and are the statistical objects we now essentially focus on.

In the Lagrangian framework, the quantities $C_1$ and $C_2$ can  be conveniently written in terms of  single-point and two-point \emph{forward} transition probabilities from the source, obtained by averaging the stochastic trajectories (\ref{eq:stochtraj}) over both the noise and the realization of the velocity field (\cite{gawedzki_stochastic_2008})~:  
\begin{equation}
	\label{eq:transitions}
	\begin{split}
		& p_1(\bx,t|t_0) := \av{p_\kappa\left(\bx,t|0,t_0\right)}, \hspace{0.1cm } \text{and}\\
		& p_2(\bx,\bx^\prime,t|t_0,t_0^\prime) := \av{ p_\kappa\left(\bx,t|0,t_0\right)p_\kappa\left(\bx^\prime,t|0,t_0^\prime\right)}.
	\end{split}
	\end{equation}
The \wit\; statistics then read 
	\begin{equation}
		\begin{split}
		& C_1(\bx,t) = 0, \hspace{0.2cm} \text{and}\\
		&C_2(\bx,\bx^\prime, t) = \phi_\sigma^2 \int_{0}^t \dt_0 \,p_2(\bx,\bx^\prime,t|t_0,t_0),
		\end{split}
	\label{eq:witstat}
	\end{equation}
while the \cit\; statistics are obtained as 
	\begin{equation}
	\begin{split}
		&C_1(\bx,t) =  \phi_0 \int_{0}^t \dt_0\, p_1(\bx,t|t_0), \hspace{0.1cm} \text{and} \\
		&C_2(\bx,\bx^\prime, t) = \phi_0^2 \int_{0}^t \dt_0 \int_{0}^t \dt_0^\prime\, p_2(\bx,\bx^\prime,t|t_0,t_0^\prime).
	\end{split}
	\label{eq:citstat}
	\end{equation}

Due to the linear dependence of the concentration field with respect to the source term, and the white-in-time nature of the \wit\, source, one obtains the average and correlation field induced by the full source (\ref{eq:realsource}) as the sum of the \wit\, and \cit\, statistics. The average concentration is solely prescribed by the \cit\, contribution, but the correlation has both non-trivial \cit\, and \wit\, contributions. Let us point out, that higher order statistics would involve additional correlations between the \wit\, and the \cit\, terms.\\

\subsection{Isotropic correlation and quasi-Lagrangian mass.}
To characterize the correlation field $C_2(\bx,\bx^\prime,t)$, it proves convenient to introduce the midpoint $ \boldsymbol{\rho} = \dfrac{\bx+\bx^\prime}{2}$ and the separation $\br = \bx -\bx^\prime$. In the steady state, the isotropic nature of the advecting velocity field makes $C_2$ only depend on three parameters~: \emph{(i)} the absolute distance $\rho = | \boldsymbol{\rho}|$, \emph{(ii)} the relative distance $r = | \br|$  and \emph{(iii)} the angle $\theta = (\br,\brho)$, so that $C_2 = C_2(r,\rho,\theta)$.

\begin{figure}
	\centering
%	\begin{minipage}{0.5\textwidth}
%	\centering
	\includegraphics[width=0.4\textwidth,trim=2cm 1cm 0.5cm 1cm,clip]{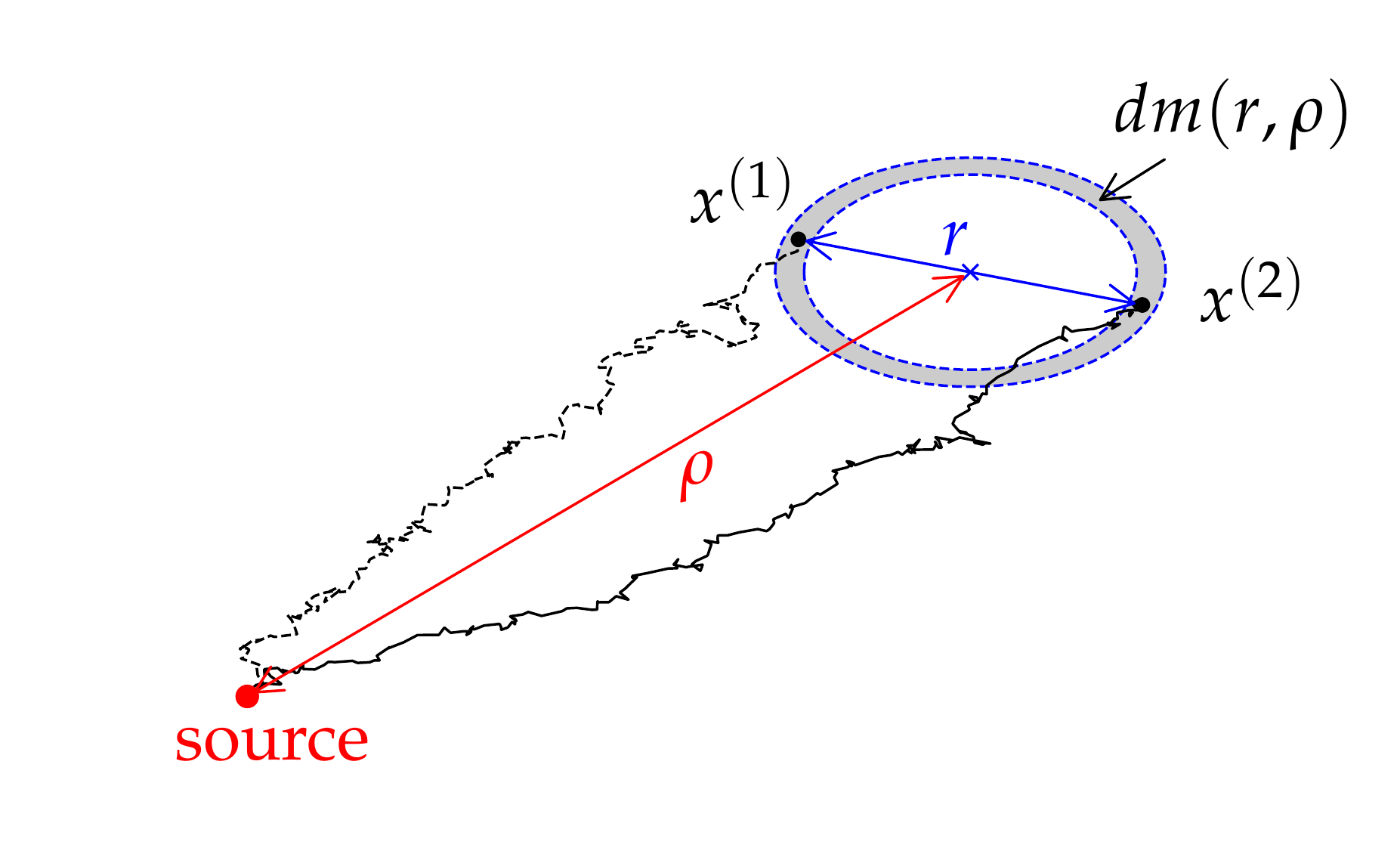}\\
	%(c) Notations
%	\end{minipage}\\
	\begin{minipage}{0.49\textwidth}
	\centering
	\includegraphics[width=\textwidth,trim=2cm 2.5cm 2cm 1.5cm,clip]{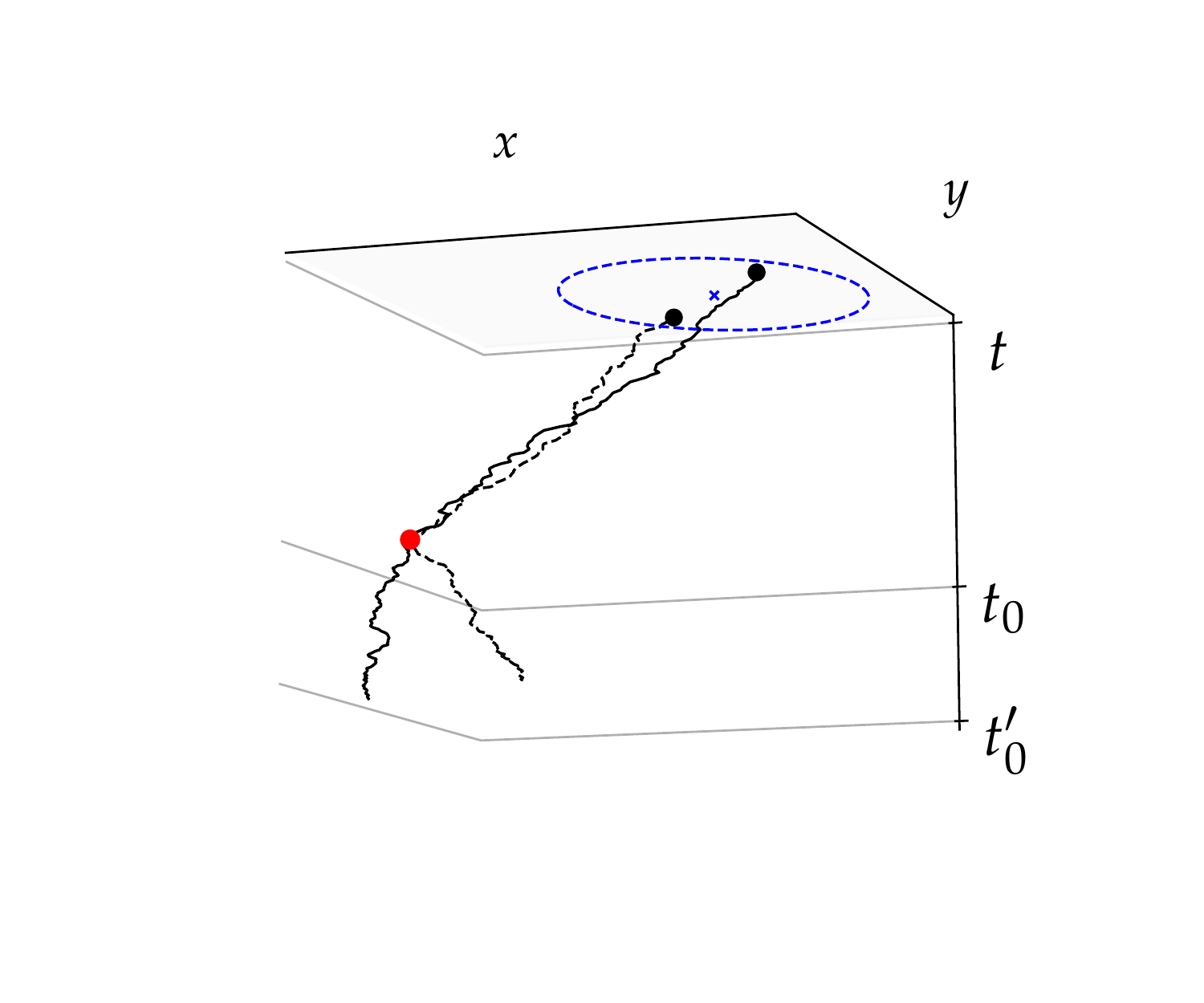}\\
	\wit\, injection
	\end{minipage}
	\begin{minipage}{0.49\textwidth}
	\centering
	\includegraphics[width=\textwidth,trim=2cm 2.5cm 2cm 1.5cm,clip]{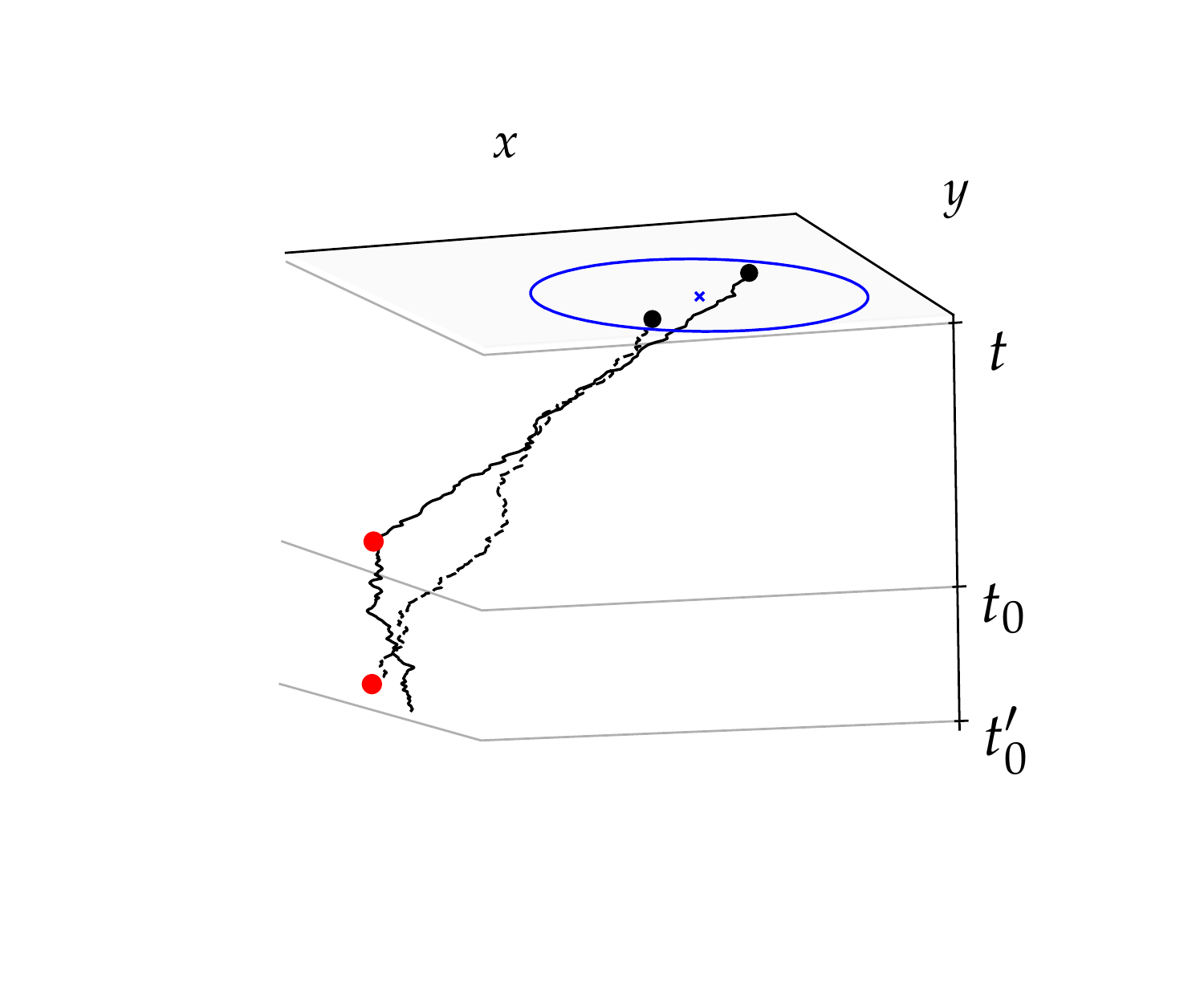}\\
	\cit\, injection
	\end{minipage}
	\caption{Qualitative distinction between \wit\,  and \cit\, injections for the Lagrangian contribution to the mass $\dd m(r,\rho)$. In the \wit\, case, correlations between tracers that transit \emph{simultaneously} through the source provide the only contribution. In the \cit\, case,  correlations between non-coincident trajectories do also contribute.}
\label{fig:sketch}
\end{figure}

To simplify the discussion, we  focus on the isotropic contribution $\mathcal c  (r,\rho)$ to the  correlation field, obtained by averaging  $C_2$ over the azimuthal solid-angle. Physically, the quantity  $\mathcal c  (r,\rho)$  relates to the concept of quasi-Lagrangian mass $d \mathcal m _{QL}$, that describes  the probability of a puff having some mass at a distance $r$, knowing that its center is located at a distance $\rho$ away from the source : 
\begin{equation}
	d \mathcal m_{QL}  (r,\rho) =r^{d-1} \dd r \mc(r,\rho)/\mZ(\rho) \hspace{0.5cm}\text{with}\hspace{0.5cm} \mZ(\rho) = \int_\epsilon^R \dd r r^{d-1}\mc(r,\rho),
\label{eq:qlmass}
\end{equation}
\modif{and where $R$ is a regularizing cut-off, that we can take to be $\infty$ if the defining integral for $\mZ(\rho)$ is convergent.}
From Equations (\ref{eq:witstat}) and (\ref{eq:citstat}), one qualitatively expects the statistics of the correlation $c(r,\rho)$ to be more intricate in the  \cit\, than in the \wit\, scenario. In the latter case, the correlations of the concentration field are due to the correlations between  pairs of trajectories that pass simultaneously trough the source. In the former case, correlations between particles  released from the source at different times do also contribute (see Figure \ref{fig:sketch}).
 To go beyond this very qualitative remark,  the space-time statistics of the turbulent velocity field  that intervenes in Equation (\ref{eq:stochtraj}) now need to be specified further.

\subsection{Kraichnan velocity ensemble.}
%\paragraph{Velocity statistics.}
In order to treat a soluble model of turbulent transport, the velocity field is now prescribed to have a vanishing correlation time and  Gaussian spatial statistics, hereby yielding statistics in the so-called ``Kraichnan velocity ensemble'' (see \cite{kraichnan_diffusion_1970,gawedzki_turbulent_2001,gawedzki_stochastic_2008}, and references therein)~:
\begin{equation}
	\begin{split}
	 &\av{v_i(\bx,t)} = 0, \hspace{0.1cm} \text{and} \\ & \av{v_i(\bx,t)v_j(\bx^{\prime},t^\prime)} = R_{ij}(\left|\bx -\bx^\prime\right|)\delta(t-t^\prime).
	\label{eq:Kraichnan}
	\end{split}
\end{equation}

The covariance matrix $R_{ij}$ is chosen so as to mimic the spatial correlations of a $d$-dimensional rough turbulent velocity field in the inertial range, taken to be homogeneous, isotropic and compressible \cite{orszag_lectures_1974,gawedzki_stochastic_2008}~: 
\begin{equation}
	\begin{split}
	& R_{ij}(r) = D_0\left( \delta_{ij} - d_{ij}(r) \right), \hspace{0.1cm} \text{where} \;\;\\
	&  d_{ij}(r) = \left(\dfrac{r}{\lambda}\right)^{\xi} \left( \gamma \delta_{ij} -\beta \dfrac{r_ir_j}{r^2}\right), \hspace{0.1cm}\text{and using}\\
	&\gamma = \dfrac{d-1+\xi(1-\cp)}{(d-1) (1+\cp \xi)},  ~\text{and}~\beta = \dfrac{\xi \left( 1- d \cp \right)}{(d-1)(1+\cp \xi)}.
	\end{split}
	\label{eq:KraichnanCov}
\end{equation}

The compressibility degree $$  \cp= \lim_{r \to 0}  \partial^2_{ij} R_{ij}(r) / \partial^2_{jj} R_{ii}(r)$$ ranges from $0$ to $1$.
The roughness of the velocity field is given by the coefficient $\xi$, which ranges from $0$ to $2$. The coefficient 
$\lambda$ represents the integral length scale, and  the inertial scales hence correspond to $r \ll\lambda$.
The parameters relevant to describe  the 3D direct cascade or the  2D inverse cascade of homogeneous isotropic incompressible turbulence are $\cp =0$ and $\xi =4/3$.

%\paragraph{Remark.}
It is well known that the Kraichnan model is not realistic, in the sense that  the statistics of a genuine turbulent velocity field are usually both non-Gaussian in space and non trivially correlated in time \cite{frisch_turbulence_1995,canet_spatiotemporal_2016}. While those features may fundamentally alter the phenomenology of tracer  dispersion \cite{chaves_lagrangian_2003}, the Kraichnan model can however be expected to provide a qualitative understanding of pure Lagrangian phenomena.
For example,  the anomalous features of passive transport have been tied in the Kraichnan model to the existence of so-called ``zero-modes'' (see for instance \cite{falkovich_particles_2001, gawedzki_turbulent_2001, gawedzki_stochastic_2008}, and references therein). The concept of zero-mode has proven fruitful for DNS, where it is reflected in terms of statistical conservation laws   \cite{pumir_geometry_2000,arad_statistical_2001,celani_statistical_2001,sreenivasan_lagrangian_2010,falkovich_introduction_2008}. %\bla{Peut etre plutot mettre le paragraphe qui precede dans l introduction}

\subsection{Steady states of the concentration field.}
Combining Equations  (\ref{eq:Kraichnan}-\ref{eq:KraichnanCov}) to Equation (\ref{eq:stochtraj}), and using standard It\^o calculus \cite{risken_fokker-planck_1984}, it is easy to show that  the transition probabilities (\ref{eq:transitions}) propagate as $\partial_t p_i = -\mathcal M_i p_i$, with
\begin{equation}
	\begin{split}
	&\mM_1[\bx] = -\left(\kappa + \dfrac{D_0}{2}\right) \partial^2_{x_ix_i} \hspace{0.1cm} \text{and}\hspace{0.2cm}\\
	& \mM_2[\bx,\bx^\prime] = \mM_1[\bx]+\mM_1[\bx^\prime]-\partial^2_{x_ix_j^\prime}R_{ij}(r).
	\label{eq:propagators}
	\end{split}
\end{equation}
It then follows from the definitions (\ref{eq:witstat}) and (\ref{eq:citstat}), that the steady statistics of the WIT concentration field satisfy ~:
\begin{equation}
	\begin{split}
		&C_1(\bx)=0, \hspace{0.1cm}\text{and}\hspace{0.1cm}\mM_2 C_2(\bx,\bx^\prime)=\phi_\sigma^2 \delta(\bx)\delta(\bx^\prime),\\
	\end{split}
	\label{eq:steady_wit}
\end{equation}
while the CIT statistics are determined by~:
\begin{equation}
	\begin{split}
		&\mM_1C_1(\bx)=\phi_0 \delta(\bx), \hspace{0.1cm}\text{and} \hspace{0,1cm}\\
		&\mM_2 C_2(\bx,\bx^\prime)=\phi_0\left( \delta(\bx) C_1(\bx^\prime) + \delta(\bx^\prime) C_1(\bx) \right).
\end{split}
	\label{eq:steady_cit}
\end{equation}

%\paragraph{The interplay between absolute and relative dispersion.}
To proceed further, it is useful to write the propagator $\mM_2[\bx,\bx^\prime]$ in terms of the absolute separation vector $\brho $ and the relative separation $\br$ as
\begin{equation*}
 \mM_2[\bx,\bx^\prime] =  \mM_1[\brho]+\dfrac{D_0}{4} d_{ij}(r)\partial^2_{\rho_i\rho_j}-\partial^2_{r_ir_j} \left(2\kappa \delta_{ij}+D_0 d_{ij}(r)\right).
\end{equation*}
As in \cite{celani_point-source_2007},  inertial range asymptotics are obtained by considering $r \ll \lambda$ and letting $\kappa \to 0$. The propagator then reduces to a sum between two operators, with one acting on the centre of mass $\brho$, and the other one on the relative separation $\br$~:
\begin{equation} 
	\begin{split}
	&\mM_2[\bx,\bx^\prime] =  \mM_1[\brho]+\mM_\xi[\br], \hspace{0.1cm}\text{with} \\
	&\hspace{0.1cm} \mM_\xi[\br] = -D_0\partial^2_{r_ir_j} d_{ij}(r).
	\end{split}
\label{eq:M2_asymp}
\end{equation}
Upon inspection of the previous equations, one qualitatively expects the behaviour of the fluctuations to depend crucially on the features of the one-point motion.
This is trivial in the  \cit\, case, where $C_1$ appears explicitly on the right-hand side of the steady-state equation (\ref{eq:steady_cit}).
For both the \wit\, and the \cit\, cases though, the propagator $\mM_2$ involves an interplay between the absolute dispersion propagator $\mM_1$ and the relative dispersion propagator $\mM_\xi$, a feature that might affect the mass statistics in a less immediate manner.
This intuition will be substantiated in the next two sections. 
\begin{comment}
In the next section, the equations (\ref{eq:steady}) and  (\ref{eq:M2_asymp}) are used both in the the \wit\, and the \cit\, cases, in order to  determine the inertial behaviour  of the isotropic contribution to the fluctuations.

Because of the specific interplay between the absolute and the relative dispersion, one anticipates the fluctuation field in the Kraichnan model  not to be faithfully representative of a ``genuine'' turbulent flow, be the latter experimental or obtained for  by direct numerical simulations of the Navier-Stokes equations.   $\mM_1$ being essentially a diffusion operator,  the center of mass $\rho$ is predicted to diffuse throughout the dispersion. This is certainly not the case in time-correlated turbulent flows, unless for time scales far  greater than  the integral time scale (\cite{taylor_diffusion_1922}). Single tracers are there  expected to be swept away by the large scales, hereby experiencing  successively a ballistic phase and a hyper-diffusive regime, an observation which has motivated in the past the development of Lagrangian stochastic models in terms of Langevin process (see for instance \cite{wilson_review_1996, lacasce_statistics_2008}).

This  issue will be illustrated in a more quantitative fashion in Section 4, where we will discuss how the concentration statistics are modified, if  $\mM_1$ is taken  to describe a ballistic rather than a diffusive motion.
\end{comment}

\section{Fluctuation statistics in the Kraichnan ensemble}
This section discusses the statistics of the concentration field in the Kraichnan model, and contrast  \wit\, and \cit\, statistics. While the effective computation is only described in outline, technical details can be found in Appendix \ref{sec:details}. 
 For the purpose of brevity,  the isotropic contribution $\mathcal c(r,\rho)$ of the correlation field  is later referred to as being itself the ``correlation field''. 

\subsection{Computing the correlation field.}
Both the \wit\, and the \cit\, correlation fields $\mc(r,\rho)$ are obtained by taking the Hankel transforms of Equations (\ref{eq:steady_wit}-\ref{eq:steady_cit}) with respect to $\rho$. 
More explicitly, we look for a solution in the form :
\begin{equation*}
	\begin{split}
	&	\mc(r,\rho) = \int_0^{\infty}\dd k \,k^{d-1} \mJ_0(k\rho) \,\mc(r,k), ~~ \text{so that}\\
	&  \mc(r,k) := \left(\dfrac{2}{\pi}\right)^{d-2}\, \int_0^{\infty}\dd \rho \,\rho^{d-1} \mJ_0(k\rho)\,\mc(r,\rho),
	\end{split}
\end{equation*}
and then solve for $\mc(r,k)$.
$\mJ_0$ is  shorthand for the standard Bessel function of the first kind  $J_0$ when $d=2$, and the spherical Bessel function of the first kind   $j_0$ when $d=3$.
The Hankel transforms of the steady equations (\ref{eq:steady_wit}-\ref{eq:steady_cit}) read~:
\begin{equation}
	\begin{split}
	&\left(M_\xi[r] + \dfrac{D_0}{2} k^{2} \right) \mc(r,k) = \rhs(r,k), \text{~~where} \\
	& \rhs(r,k) =
%	\begin{split}
		\dfrac{\phi_\sigma^2}{2\pi^{d-1}} \delta_\epsilon(r) \hspace{0.2cm}\text{(\wit \,case)}, \\		
	\text{or~~}&\rhs(r,k) =
		\dfrac{\phi_0}{\pi^{d-1}} \mJ_0\left( \dfrac{kr}{2}\right) \mc_1(r) \hspace{0.2cm} \text{(\cit \,case).}	
%	\end{split}
	\end{split}
	\label{eq:hankelsteady}
\end{equation}
\modif{ To deal carefully with the $\delta$ function involved in the \wit\, right-hand side, the small source extension $\epsilon$ is here again made explicit.}%, hence the $\epsilon$ subscript present  in the expression for $\rhs(r,k)$ in the \wit\, case (see  Paragraph \ref{sec:HankelDetails} for details).
The coefficients $\mc(r,k)$ are then found explicitly after some long but straightforward algebra. Reconstructing the correlation field there from yields the final  result~:
\begin{widetext}
\begin{equation}
	\begin{split}
	&\mc(r,\rho)=	2^{1-d/2}\, \left(2-\xi \right)^{d-1} \,D_0^{-1}\, \lambda^{\xi(1-d/2)} r^{(d/2-1)(\xi-2)} \left(c_-(r,\rho)+c_+(r,\rho)\right), \hspace{0.1cm} \text{with}\\
	%\text{with~}
	 &\mc_-(r,\rho)=	 %r^{(d-2)(\xi/2-1)} 
\int_0^\infty \dd u\, u^{d-1}\mJ_0\left(us\right) K_\omega\left(u\right) \int_0^1 \dd v v^{\frac{1+m}{2} -\xi} \rhs\left(rv,\dfrac{u s}{\rho}\right) I_\omega\left(uv^{1-\xi/2}\right), \\
\text{and}	\hspace{0.2cm} %\text{and~}
	&\mc_+(r,\rho)=	 %r^{(d-2)(\xi/2-1)}
 \int_0^\infty \dd u\, u^{d-1}\mJ_0\left(u s\right) I_\omega\left(u\right) \int_1^\infty \dd v v^{\frac{1+m}{2} -\xi} \rhs\left(rv,\dfrac{u s}{\rho}\right) K_\omega\left(uv^{1-\xi/2}\right).
	\end{split}
\label{eq:general}
\end{equation}
\end{widetext}
\modif{$I_\omega$ and $K_\omega $are the modified Bessel functions of the first and second kind order $\omega$.}
The previous formula involves a crucial dimensionless parameter, that will be later commented on : 
\begin{equation}
	s^2 := \dfrac{(2-\xi)^2}{2} \left(\dfrac{\rho}{r}\right)^2\left(\dfrac{r}{\lambda}\right)^{\xi},
\end{equation}
along with the explicit coefficients
\begin{equation}
\begin{split}
& \omega:= \dfrac{f}{2-\xi} \hspace{0.3cm} \text{where~} f= \left( (m-1)^2 -n\right) ^{1/2},\\
&
m := (d+\xi-1)\left( 1+\frac{\cp \xi}{1+\cp \xi}\right), 
\hspace{0.1cm}\text{and}~~\\
& n := (d+\xi-2)\left( d+\xi \right) \dfrac{\cp \xi}{1+\cp \xi}.
\end{split}
\label{eq:coeff}
\end{equation}
Plugging the expressions (\ref{eq:hankelsteady}) into  the general expression (\ref{eq:general}) yields the final result. Quite remarkably, the integrals (\ref{eq:general})  can be computed explicitly in the \wit\,  case. While this is not so in the \cit\, case, asymptotic scaling regimes can still be identified.
 \subsection{\wit\, statistics}
\subsubsection{\wit\, Fluctuations.}
The explicit expression for the \wit\, correlations is :
%\begin{equation}
%	\mc(r,\rho) =  \drac{\phi_0^2}{D_0} \dfrac{\epsilon^g}{g+d} \dfrac{r^{}}{}
%\end{equation}
%
%
\begin{equation}
	\begin{split}
	&\mc(r,\rho) = c_d
  \left(\dfrac{\epsilon}{r}\right)^g%
\left(\dfrac{r}{\lambda}\right)^{\xi(d/2-1)}
\dfrac{r^{2-2d}}{\left(1+s^2\right)^{\omega+d/2}}, \hspace{0.1cm}\text{with}\\%
	& c_d =  \dfrac{2^{d/2}\Gamma\left(\omega+\frac{d}{2}\right)}{\pi^{d/2}V_d \,\Gamma\left(\omega+1\right)}%
\dfrac{(2-\xi)^{d-1} }{g+d}\dfrac{\phi_\sigma^2}{D_0}, 
	\end{split}
\label{eq:witfluct}
\end{equation}
and where  $g := \dfrac{3+m+f}{2}-\xi-d$, $V_d$ is the $d$-dimensional volume of the unit-sphere, and the other coefficients given by (\ref{eq:coeff}).
The scaling behaviors  of the correlation depend on the large-scale $\lambda$ and are therefore anomalous.
As in the incompressible case discussed by \cite{celani_point-source_2007}, the specific scaling properties are controlled by the  value of the dimensionless coefficient $s^2$. This coefficient  is essentially a ratio between two timescales, namely $s^2 \sim \frac{ \tau_0(\rho)}{\tau_\parallel(r)}$,  where $\tau_0(\rho) \sim \frac{\rho^2}{D_0}$  and $\tau_\parallel(r) \sim \frac{r^2}{D_0 (r/\lambda)^\xi}$ respectively represent  the Lagrangian time-scales for the absolute and relative separations. 
In those asymptotics, and without keeping track of the constants, the correlation behaves as~:
\begin{equation}
\mc(r,\rho) \sim
	\begin{cases}
	&\left(\dfrac{r}{\epsilon}\right)^{-g} \left(\dfrac{r}{\lambda}\right)^{\xi(d/2-1)}r^{2-2d} \hspace{0.3cm}\text{for $s \ll 1$}\\
	&\left(\dfrac{r}{\epsilon}\right)^{-g} \left(\dfrac{r}{\lambda}\right)^{-\xi(1+\omega)} \left(\dfrac{r}{\rho}\right)^{2 \omega+d} r^{2-2d} \hspace{0.3cm}\text{for $s \gg 1$}.
	\end{cases}
	\label{wit_asymp}
\end{equation}

\modif{Let us here emphasize that while the ratio $s^2$ is a ratio between the two Lagrangian quantities $\tau_0(\rho)$ and $\tau_\parallel(r)$, it here intervenes as a parameter for the stationary {\emph{ Eulerian} field} $\mc_2(r,\rho)$. Upon suitable normalization, the field $\mc_2$  can be thought of as the \emph{Eulerian  probability} that in the stationary sate, the mass present in the domain lies in some  infinitesimal shell volume $r^{d-1}dr$ centered around a position at a distance $\rho$ away from the source (see Figure \ref{fig:sketchscaling}). All values of $s$ are therefore allowed. Shells characterized by $s^2 \simeq 1$ are those for which the Eulerian probability field is determined by the typical Lagrangian events, namely the bulk of Richardson's distribution.  Similarly, shells with a small spatial extension are characterized by small values of $s^2$, and the Eulerian probability is then determined by those particles that separate faster than average. Conversely, large values of $s^2$ relate to untypical trajectories that do not separate. In the Lagrangian framework, those  would correspond to the left-end tail of Richardson's distribution.}

\subsubsection{Effects of compressibility.}
\begin{figure}
\includegraphics[width=0.49\textwidth,trim=0cm 0cm 0cm 0cm,clip]{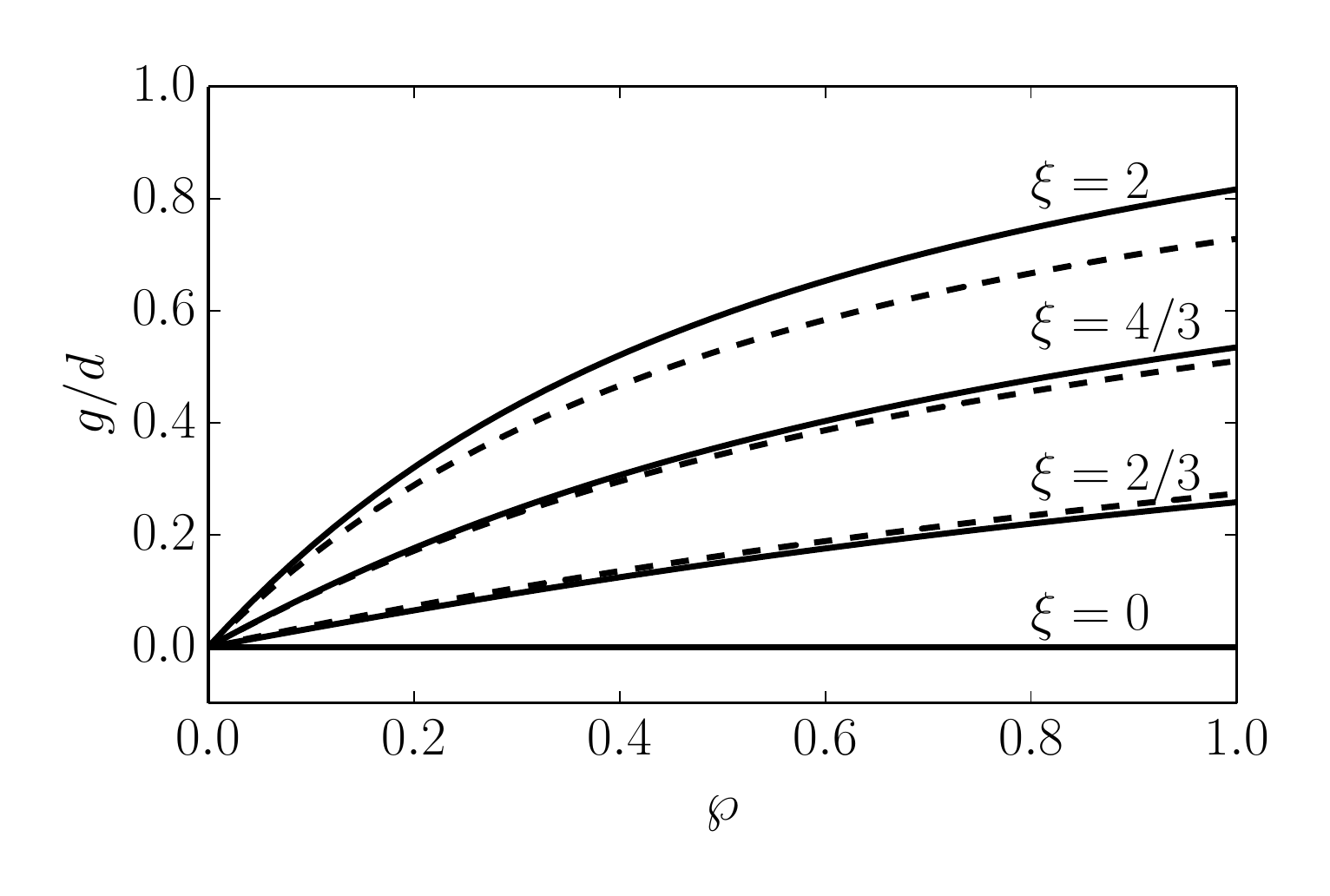}
\includegraphics[width=0.49\textwidth,trim=0cm 0cm 0cm 0cm,clip]{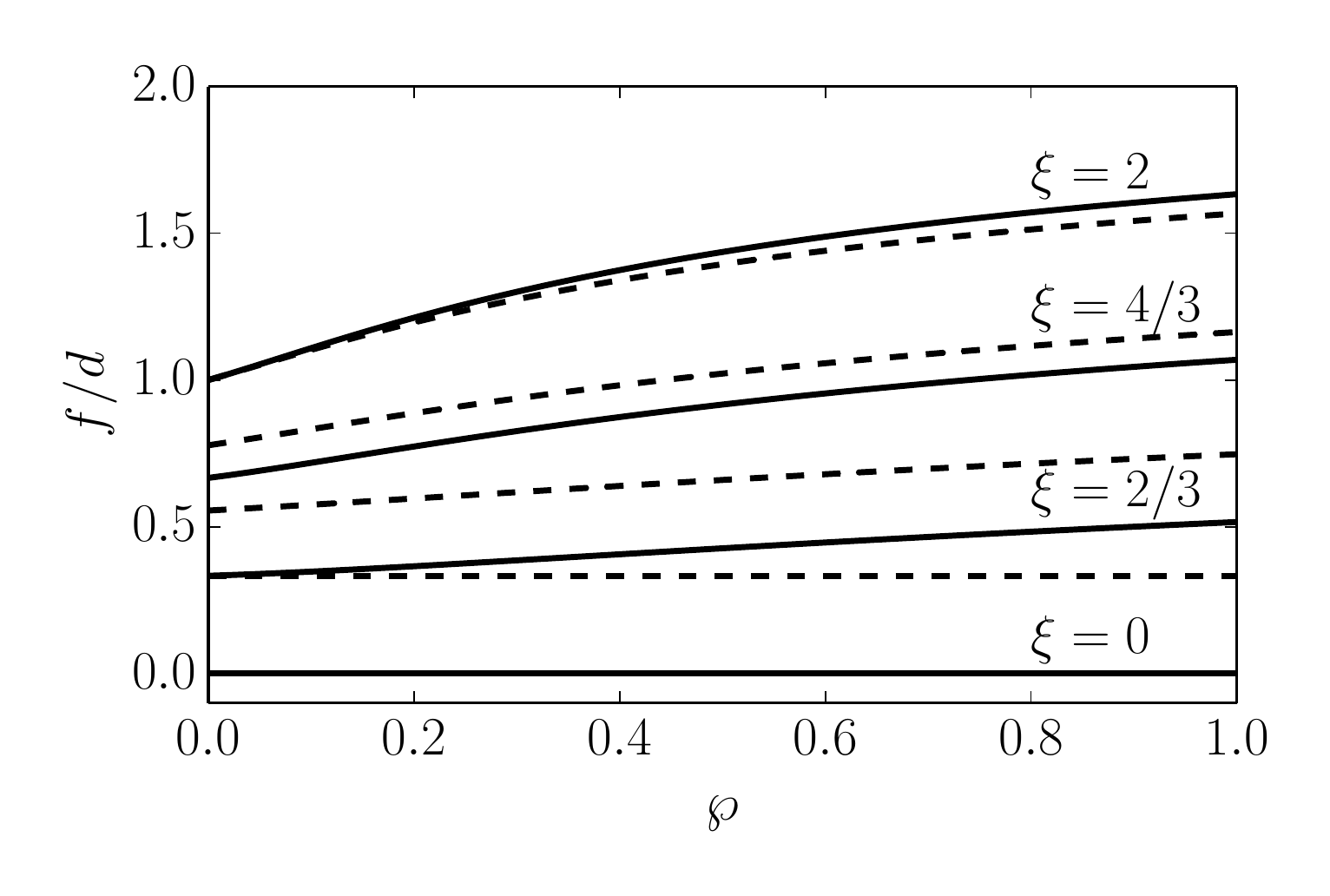}
\caption{Dependence of the exponents $g$ and $f = (2-\xi) \omega$ with respect to the compressibility degree $\cp$ for various values of the roughness coefficient $\xi$, and $d=2$ (solid) or $d=3$ (dashed).}
\label{fig:compressibility}
\end{figure}

\begin{figure}
\begin{minipage}{\textwidth}
\includegraphics[width=0.49\textwidth,trim=0cm 0cm 0cm 0cm,clip]{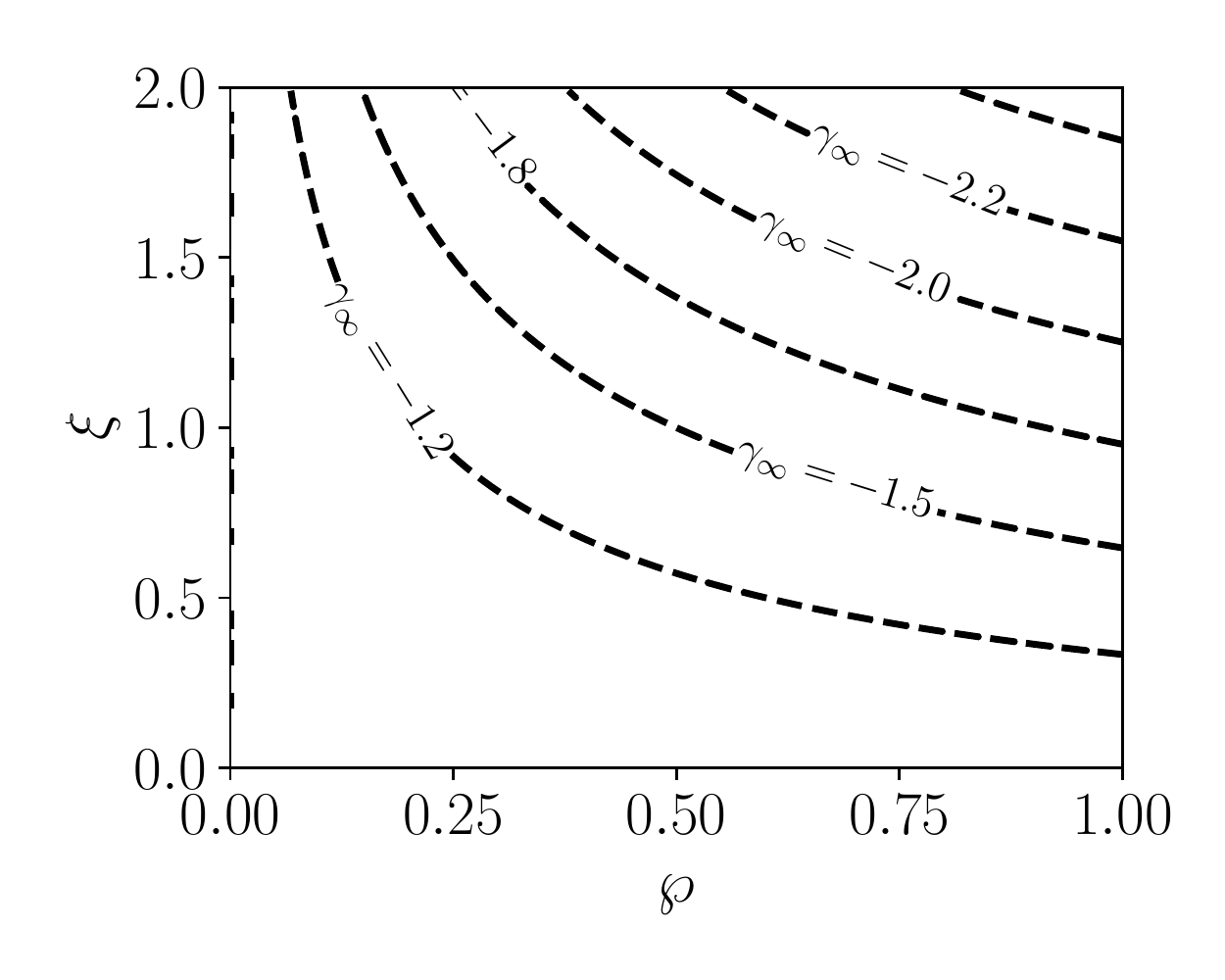}
\includegraphics[width=0.49\textwidth,trim=0cm 0cm 0cm 0cm,clip]{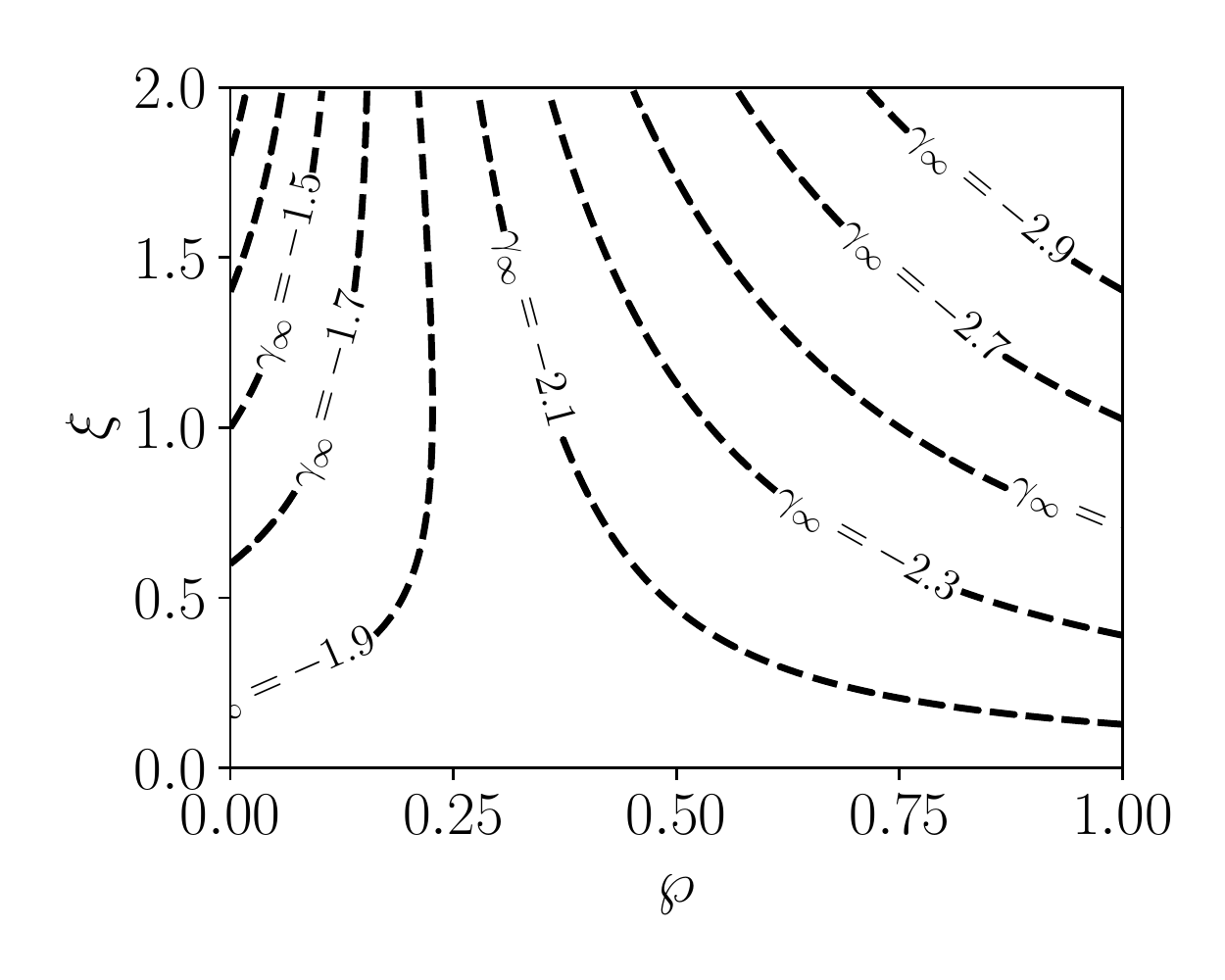}\\
\includegraphics[width=0.49\textwidth,trim=0cm 0cm 0cm 0cm,clip]{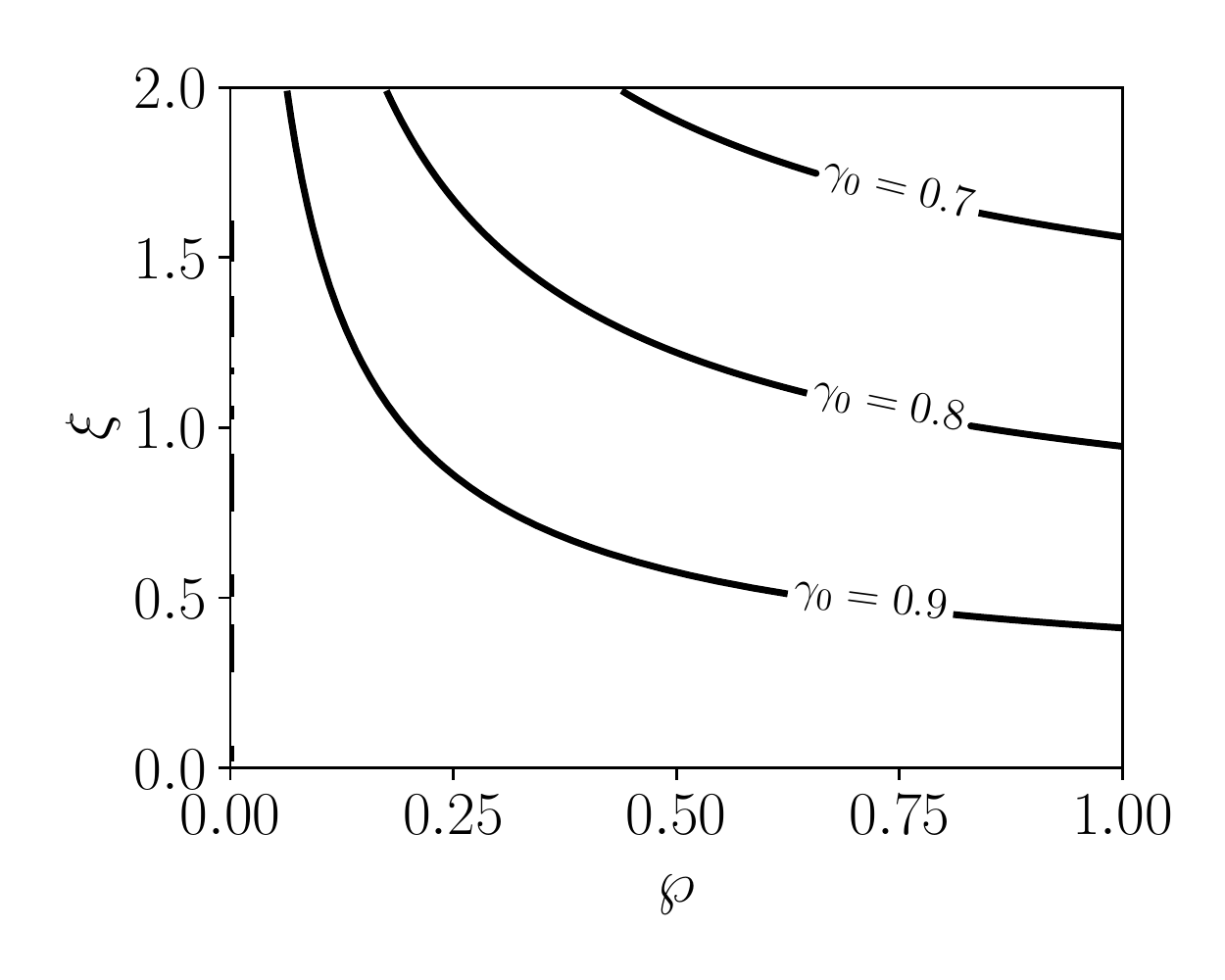}
\includegraphics[width=0.49\textwidth,trim=0cm 0cm 0cm 0cm,clip]{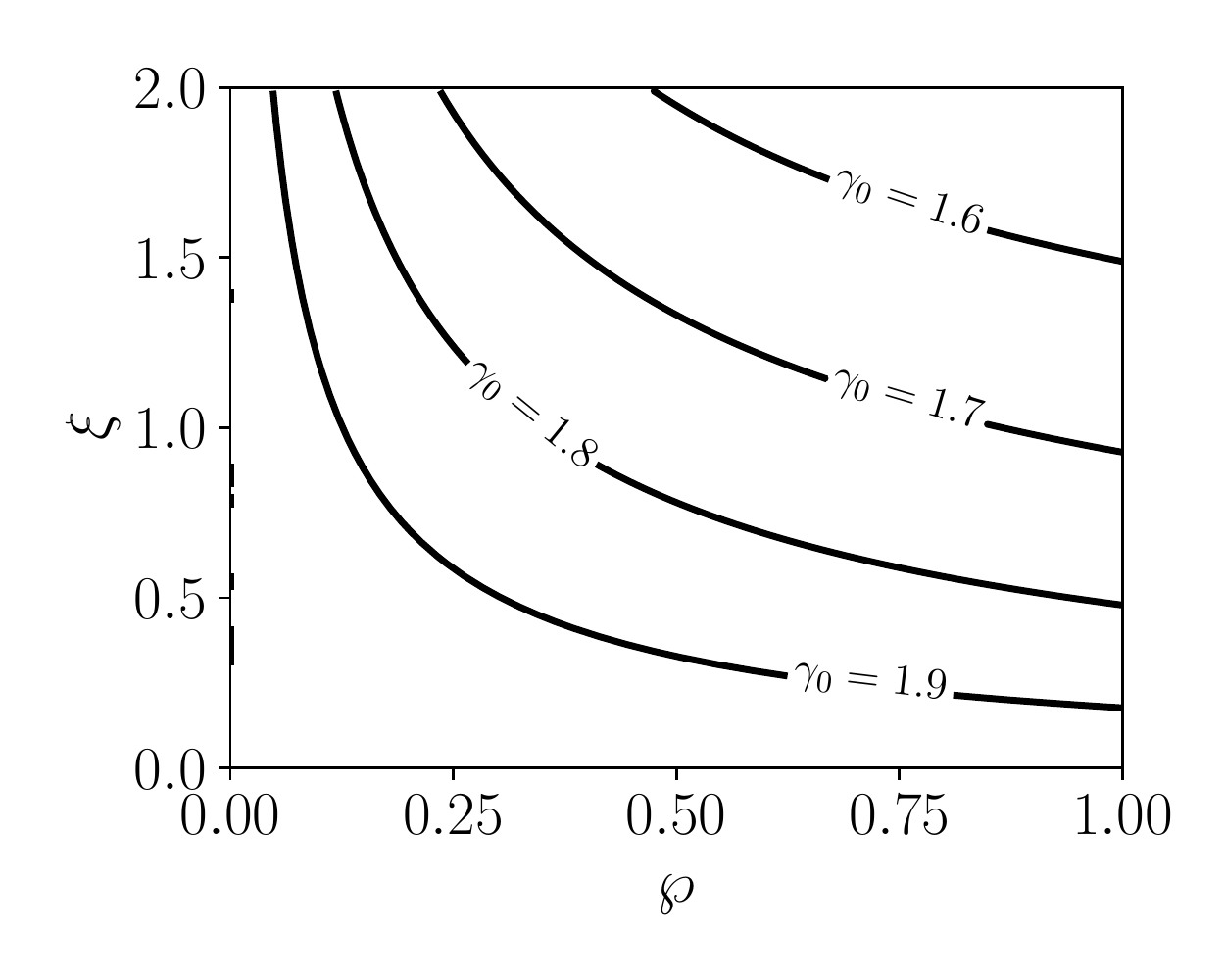}
\end{minipage}
\caption{Scaling exponents $\gamma_\infty$ (top) and $\gamma_0$ (bottom) defining the large-$r$ and small-$r$ dependence of the quasi-Lagrangian mass  $ d \mathcal  m_{QL} \sim r^{\gamma} d r $, for $d=2$ (left side) and 
$d=3$ (right side) for  the \wit\,Kraichnan case.}
\label{fig:compressibility4QL}
\end{figure}

\modif{
Let us first observe  that while a non-vanishing  compressibility seemingly only mildly affects the scaling exponents (see Figure \ref{fig:compressibility}), it makes the point-source problem become degenerate in the limit of an infinitesimal source extension $\epsilon \to 0$. Because the coefficient $g$ is strictly positive unless the underlying flow is incompressible (see the left panel of Figure \ref{fig:compressibility}), the correlation should vanish in that limit. }
\modif{
One way to circumvent the problem and define a non-trivial limit $\epsilon \to 0$  is to focus on the properties of the quasi-Lagrangian mass $d \mathcal m_{QL}(r,\rho) = \dfrac{1}{\mZ(\rho)} r^{d-1}  \mc(r,\rho) d r$  for $\rho$ strictly positive, with the normalization $ \mZ(\rho) = \int_0^\infty r^{d-1}  \mc(r,\rho) d r$. With this choice of  normalization, the quasi-Lagrangian mass becomes independent of  the source extension $\epsilon$, that can safely be taken to $0$.
This definition naturally relies upon the quasi-Lagrangian mass being indeed integrable as $r\to 0$ and $r\to \infty$, and hence on the exponents  $\dsp \gamma_{0} := \lim_{r \to 0} \dfrac{\log  d \mathcal m_{QL} /dr}{\log(r)} >-1$ and $\dsp \gamma_{\infty} := \lim_{r \to \infty} \dfrac{\log  d \mathcal m_{QL} /dr}{\log(r)}<-1$. Figure \ref{fig:compressibility4QL} shows that this is indeed the case unless  $d=2$ and the flow is incompressible. Only in  that specific case, does  one need to introduce a large-scale cut-off $R$ in the definition of $ \mZ(\rho)$. }\\

\modif{
It is known from previous work that, Lagrangian trajectories advected by a compressible Kraichnan velocity field are essentially explosive for small values of the compressibility and become ``sticky'' when the compressibility increases above the critical value $d/\xi^2$ \cite{gawedzki_sticky_2004}. The  phenomenon can be qualitatively related to the presence of shocks. In our point-source setting,  compressibility shapes the  scaling behavior of the quasi-Lagrangian mass (\ref{eq:qlmass}). A phase transition can be identified by looking at the statistics of the large shells, which are characterized by the exponent $ \gamma_\infty$.  
The top panel of Figure \ref{fig:compressibility4QL} shows an apparent transition at the critical value $\wp^\star$ defined by $\partial_\xi \gamma_\infty =0$, whereby the exponent $\gamma_\infty$ become independent of the roughness of the velocity field. We identify $\wp^\star = 0$ for $d=2$ and $\wp^\star \simeq 0.25$ for $d=3$. For $\wp>\wp^\star$,  $\gamma_\infty$ decreases  with $ \xi$~:  The  mass distribution becomes steeper as smoothness increases ($\xi \to 2$). In other words, the sticky behavior due to compressibility prevents the mass to spread broadly, unless the flow is very rough ($\xi \to 0$). The critical value of $ \wp^\star$ can be interpreted as  a case where the compressive ``stickiness'' compensate the roughness-induced explosive behavior. %This is slightly counter-intuitive as one would expect that compressibility would on the contrary increase the amount of particles present in the domain. Yet, it also probably prevents the particles from having recurrent trajectories, hence allowing for a stationary state without a cut-off. %consistent with the remark that in this case a cut-off is necessary to reach a steady state for the concentration field.
}\\

Let us finally  note that  Formulas (3.2) and (4.3) of Reference \cite{celani_point-source_2007}  are recovered as special cases of (\ref{eq:witfluct}) and (\ref{wit_asymp}), in the  incompressible limit $\cp \to 0$.

%Quite surprisingly, no transition from explosive to sticky  behaviour is found as would be expected from previous results on relative dispersion in the compressible Kraichnan model \cite{gawedzki_sticky_2004}. This suggests that Formula (\ref{wit_asymp}) may in fact only be valid for small values of $\cp$. Another possibility may be that in order to get a well-defined limit, the injection rate of the correlation $\phi_0$ has to be scaled accordingly with the source size $\epsilon$.
%Finally, note that  Formulas (3.2) and (4.3) of Reference \cite{celani_point-source_2007}  are recovered as special cases of (\ref{eq:witfluct}) and (\ref{wit_asymp}), in the  incompressible limit $\cp \to 0$.
%
%\bla{Discuss the effect ot compressibility and the sign of $g$ and $f$. Show some plots.}
%
%
 \subsection{\cit\, statistics}
\subsubsection{\cit \,average concentrations.}
Computing the \cit\, fluctuation field requires to know the \cit\, average concentration  $C_1(\bx) =: \mc_1(|\bx|)$. The latter   is obtained by direct integration  of (\ref{eq:steady_cit}), with the prescription that  $\mc_1$ vanishes at  $\infty$. For $d=2$, this is only possible provided a large scale cut-off $L$ is introduced \footnote{This feature is due to the recurring nature of the Brownian motion for $d=2$, and its transiting nature for $d=3$.}, so that
\begin{equation}
	\mc_1(|\bx|) = 
	\begin{cases}
		&-\dfrac{\phi_0}{\pi D_0} \log \dfrac{ |\bx|}{L} H(L-|\bx|) \hspace{0.2cm} \text{for $d=2$},\\
 		&  \dfrac{\phi_0}{2\pi D_0 |\bx|}   \hspace{0.2cm} \text{for $d=3$},
	\end{cases}
\end{equation}
where $H$ here denotes  the Heaviside function, that takes value $1$ for positive arguments and vanishes  otherwise.
\subsubsection{\cit \,correlations.}
 While the  \cit\, correlation field does not seem to have a fully explicit expression beyond (\ref{eq:general}), scaling behaviours can still be analysed. From (\ref{eq:hankelsteady}),  one computes  $\rhs\left(rv,u s/\rho\right) = \phi_0\pi^{1-d}\mc_1(rv)\,\mJ_0(uv s r/2\rho)$, and observes that the integrand quantities of  (\ref{eq:general}) now depend not only on the coefficient $s$ but also on the value $\Lambda := r/\rho$. 
Using the asymptotic properties of the modified Bessel functions $I_\omega$ and $K_\omega$ \cite[Chapter 9]{abramowitz_handbook_1964}, three different asymptotic regimes can be explicitly determined~: \emph{(i)} $s \gg 1$, \emph{(ii)} $s\ll1 ~\text{and}~\Lambda \gg1$ ,  \emph{(iii)} $s\ll 1~ \text{and}~ \Lambda \ll 1$, as explained in details in Section \ref{eq:citdetails}.
For each of those three regions, one can identify the following behaviors (the constants are here documented)~:
\begin{itemize}
\item\emph{(i)} $s \gg 1$ :
\begin{equation}
	\begin{split}
&\mc(r,\rho) \sim \dfrac{2^{2-d}}{\omega \pi(2-\xi)} \left(\dfrac{r}{\lambda}\right)^{-\xi} \left(\dfrac{r}{2 \rho}\right)^{\xi-\frac{m \pm f + 3 }{2}}\hspace{-0.8cm}%
\mc_1(2 \rho ) \dfrac{\phi_0}{D_0}r^2\rho^{-d}, \\
&\hspace{0.5cm} \text{with}~ \pm = \text{sign}({r-2\rho}).
	\end{split}
\label{eq:cit_larges}
\end{equation}
\item\emph{(ii)} $s\ll1 ~\text{and}~\Lambda \gg1$ :
\begin{equation}
\mc(r,\rho) \sim \dfrac{2^{5(d-2)/2}}{\pi^{d-2}} %
\left(\dfrac{r}{\lambda}\right)^{\xi}%
\mc_1(r ) \dfrac{\phi_0}{D_0}r^{2-d}\,\, \chi_d(s)
% \left(\dfrac{r}{2 \rho}\right)^{d+\xi-\frac{m \pm f + 3 }{2}}\hspace{-0.3cm}%
%\mc_1(r ) \dfrac{\phi_0}{D_0}r^{2-d}\,\, \chi_d(s) \text{~where~} \chi_d(s)=
%\begin{cases}
%	&-\log s~ \text{for $d=2$}\\
%	&\dfrac{\pi}{2}~\text{for $d=3$}
%\end{cases}.
\label{eq:cit_smallslargeL}
\end{equation}
\item\emph{(iii)} $s\ll1 ~\text{and}~\Lambda \ll1$ :
\begin{equation}
\mc(r,\rho) \sim \dfrac{2^{3(d-2)/2}}{\pi^{d-2}} %
\left(\dfrac{r}{\lambda}\right)^{\xi}%
% \hspace{-0.3cm}%
\mc_1(r ) \dfrac{\phi_0}{D_0}\rho^{2-d}\,\, \chi_d(s),% \text{~where~} \chi_d(s)=
\label{eq:cit_smallssmallL}
\end{equation}
where  in the last two formulas $ \chi_d(s)=-\log s$ for $d=2$
and  constant $ \dfrac{\pi}{2}$ for $d=3$.
\end{itemize}
%
%\begin{figure}
%\includegraphics[width=0.49\textwidth,trim=0cm 0cm 0cm 0cm,clip]{cit_plus}
%\includegraphics[width=0.49\textwidth,trim=0cm 0cm 0cm 0cm,clip]{cit_minus}
%\caption{The coefficients $\xi-(m\pm f+3)/2$ appearing in the $s\gg1 $ \cit\, scaling region,  plotted with respect to the %compressibility degree $\cp$ for various values of the roughness coefficient $\xi$, and $d=2$ (solid) or $d=3$ (dashed).}
%\label{eq:cit comp}
%\end{figure}
µ
\begin{figure}
\includegraphics[width=0.49\textwidth,trim=0cm 0cm 0cm 0cm,clip]{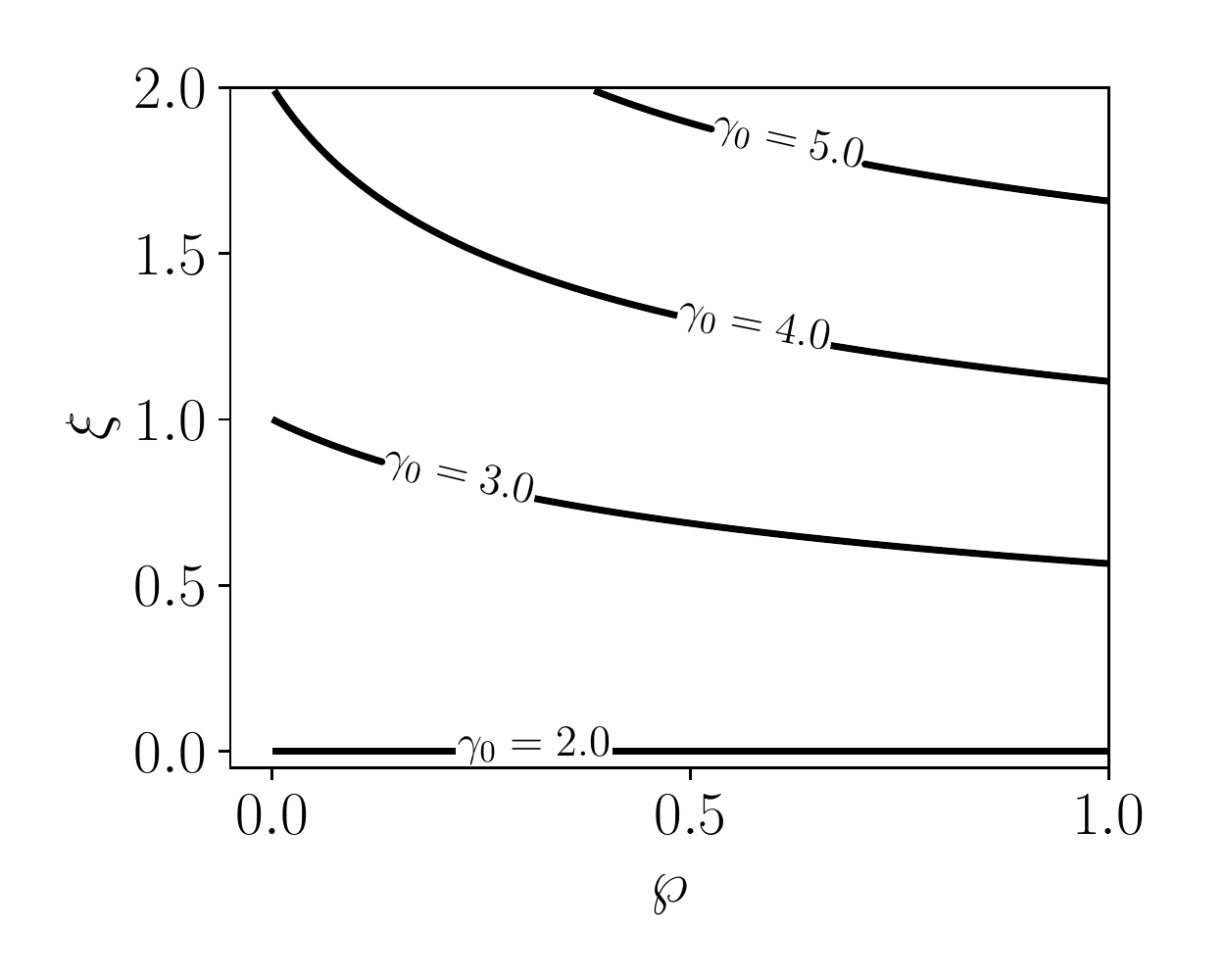}
\includegraphics[width=0.49\textwidth,trim=0cm 0cm 0cm 0cm,clip]{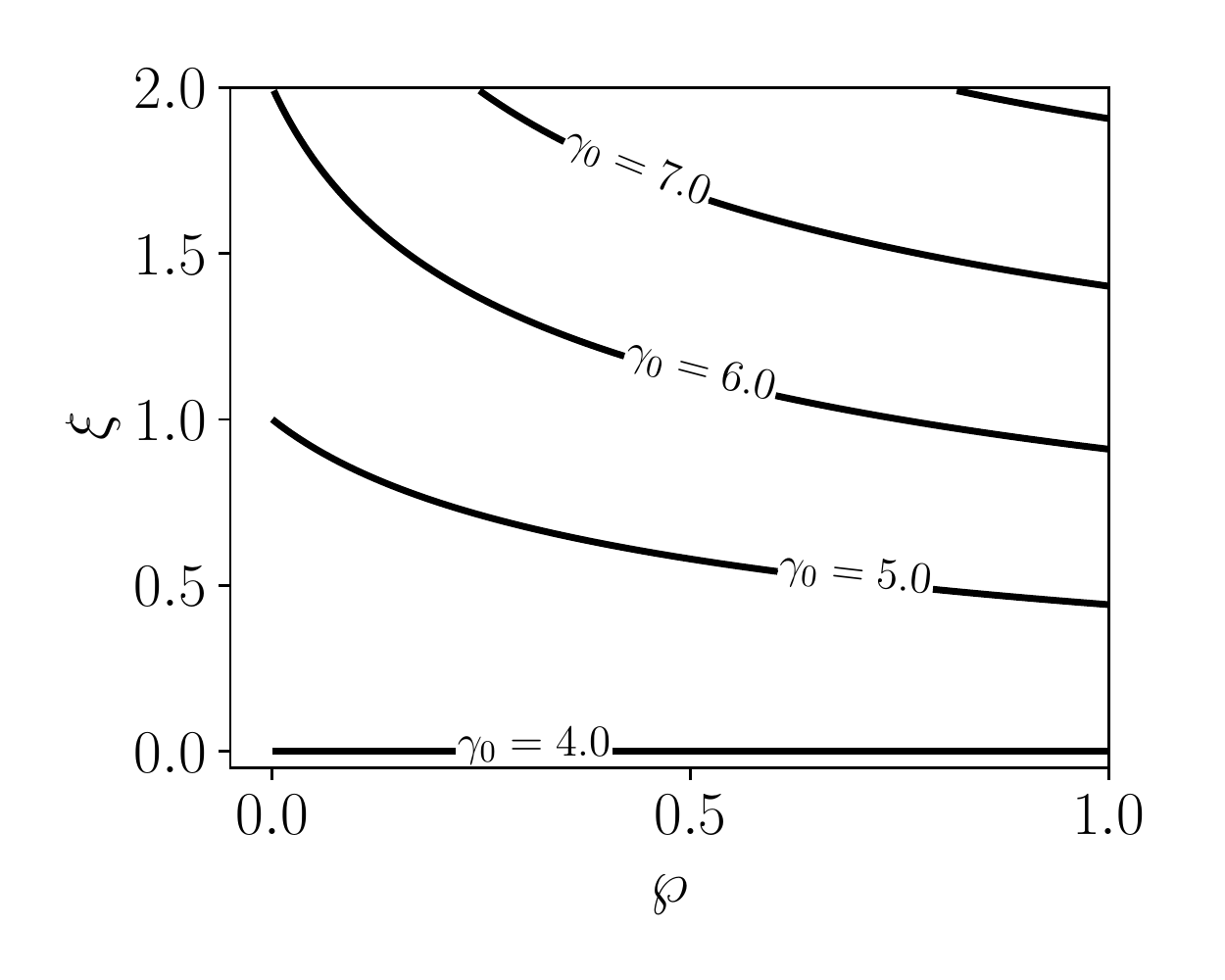}
\caption{The small $r$ scaling exponent $\gamma_0 = d+(f-m-1)/2$ of the quasi-Lagrangian mass in the case of a \cit\, injection, for $d=2$(left) and $d=3 $ (right).}
\label{eq:cit comp}
\end{figure}
%
%\paragraph{Comments.}
Let us first  observe, that while in 2D, logarithmic corrections are present  far from the source ($s\ll 1$),  scaling regions can be identified in all three asymptotic regimes, in spite of the strong Lagrangian mixing due to the continuous nature of the injection. As in the \wit\, case the scaling is intermittent, in the sense that it is affected by the integral scale $\lambda$.
 The second observation is that the asymptotic behaviors  here depend on both the timescale ratio and the aspect ratio  $\Lambda = r/ \rho$ between the relative and absolute dispersion.
\modif{
The dependence on $\Lambda$ is not surprising~: averages over shells that incorporate the source correspond to $\Lambda \ge 2$, and  incorporate the constant contribution of the  injection rate, a feature that can naturally be expected to alter the  statistics. The dependence on the Lagrangian time-scales via the coefficient $s^2$ is however more surprising, as one could have expected that this ratio was tied to the fact that the contributing Lagrangian trajectories  would pass simultaneously through the source. The present results show that this is however not the case.
} \\

Note that the compressibility degree does not here affect the statistics of the fluctuation field in a spurious way~: unlike in the \wit\, case, taking $\epsilon \to 0$ does not make the correlation field become infinite. To analyze further the  effect of compressibility,  it however remains instructive to comment  on the scaling properties of the \cit\, Lagrangian mass $ d \mathcal m \sim r^{d-1} \mc(r,\rho) d r$ , through the small and large $r$ scaling exponent $\gamma_0 $, $\gamma_\infty$, defined such that  $\dsp  d \mathcal m \underset{r\to 0 \text{\,or\,}\infty}{\sim} r^{\gamma_{0,\infty}} dr $.
Computing $\gamma_\infty = 4-d + \xi$ , it is apparent that the large-$r$ behavior is independent from the compressibility degree. This results probably owes to the fact that the continuous contribution from the source there dominates the statistics. Besides, it also shows that a large-scale cut-off $R$ needs to be prescribed for the mass to be accurately normalized.
Figure \ref{eq:cit comp} shows the iso-lines of the exponent $\gamma_0 = d+(f-m-1)/2$ for the small $r$ behavior. As in the \wit\, case, compressibility only  weakly alters the small-$r$ scaling.
% While it affects the scaling exponent in the region $s \gg 1 $ corresponding to spatial averages over small values of $r$'s (see Figure \ref{eq:cit comp}),  those exponents are found to be independent of $\cp$ for $s\ll 1$.

\section{Concentration statistics in the presence of a large-scale sweeping}
\subsection{Modeling the large-scale sweeping }
As mentioned in the introduction, the statistics of the correlation depend on the interplay between the absolute and the relative dispersion. This is particularly obvious in the \cit\, case, where the average $c_1$ appears explicitly in the expression for the correlation given by Equations (\ref{eq:cit_larges}-\ref{eq:cit_smallssmallL}). However, the specific  interplay that appears in the Kraichnan ensemble, between an absolute diffusive dispersion and a relative explosive separation can look paradoxical with respect to the Lagrangian phenomenology  of  time-correlated turbulence \emph{\`a la} Kolmogorov.
In  DNS and experiments,  both the absolute and relative separation are known to become diffusive only  at times greater than the Lagrangian integral time-scale $\tau_L$.  Below $\tau_L$, the phenomenologies of absolute and relative dispersion differ. 
On the one hand,  it is known from state-of-the art numerics that \modif{after an initial transient ballistic regime} the bulk statistics of relative separation are reasonably well-described by Richardson diffusion both in two and three dimensions,  (see \cite{boffetta_pair_2000,boffetta_relative_2002,bitane_time_2012,thalabard_turbulent_2014,bourgoin_turbulent_2015}):~ This therefore justifies the use of the relative dispersion operator $\mM_\xi$.  On the other hand, the Lagrangian velocity measured along a single trajectory is typically correlated over the integral time-scale (\cite{yeung_lagrangian_1989,mordant_measurement_2001}). Unlike the Kraichnan model, the absolute dispersion is therefore not diffusive, except for timescales far  greater than  the integral time scale (\cite{taylor_diffusion_1922}). Refined treatments of single-particle dispersion have motivated in the past the development of Lagrangian stochastic models in terms of Langevin process (see for instance \cite{wilson_review_1996, lacasce_statistics_2008}) but go beyond the point of this paper. For the present purpose, it is probably reasonable to consider that the absolute dispersion is essentially ballistic for times below $\tau_L$.

In order to investigate quantitatively how a ``Ballistic/Explosive'' interplay differs from the ``Diffusive/Explosive'' interplay studied in the previous section, the Kraichnan velocity ensemble (\ref{eq:Kraichnan}) is now altered into~:
\begin{equation}
\begin{split}
	& v_i(\bx,t) = U_0 \dfrac{x_i}{|\bx|} + u_i(\bx,t), \\
\end{split}
\label{eq:Kraichnanpuff}
\end{equation}
where $U_0$ is a constant  velocity, and  $u$ is a fluctuating turbulent field  with Kraichnan statistics, as prescribed by  Equation (\ref{eq:Kraichnan}). The ratio $D_0/U_0$ defines a length scale $r_0$, which is here assumed to be small compared to the integral length scale. \modif{Below $r_0$, the diffusive nature of the Kraichnan model dominates the one point motions, and those therefore diffuse.  Only for $r>r_0$ does the motion become ballistic. For our present purpose, we therefore wish to to consider statistics on scales $r \gg r_0$.}
 In the spirit of the so-called ``puff-particles models'' described in \cite{haan_puffparticle_1995} in the context of atmospheric dispersion modeling, the idea of Model (\ref{eq:Kraichnanpuff}) is to  prescribe the \modif{barycenter of puffs of tracers}  to have a dynamics independent from  the  fluctuating turbulent field. \modif{In the present case, the barycenter is essentially prescribed by the large-scale velocity $U_0$ and is insensitive to Kraichnan diffusion when it varies on scales greater than $r_0$. }

In the limit of vanishing diffusivity $\kappa$, one can check that the single point motion for the inertial scales  is essentially ballistic,
that is  $\av{|\bx|} = U_0 t$ for  $r_0 \ll |\bx| \ll \lambda$.  The relative motion is left unaltered and given by Kraichnan statistics. In other words, puffs of tracers move ballistically on average but spread explosively. The velocity ensemble (\ref{eq:Kraichnanpuff}) is therefore later referred to as the ``Ballistic/Explosive '' (\puff) model.
It is easily checked that the steady states equations  (\ref{eq:propagators}-\ref{eq:M2_asymp}) for the concentration statistics carry through, with the only difference that the single-point propagator is now given by~:
\begin{equation}
\mM_1^{\ball}[\bx]:= U_0 \partial_{x_i}\left(\dfrac{x_i}{|\bx|} \cdot\right).
\label{eq:m1ball}
\end{equation}

\subsection{\puff\, correlation field}
The (isotropic) correlation field of the \puff\, ensemble (\ref{eq:Kraichnanpuff}) is  computed  along the same lines as  in the previous Section, except that Laplace rather than Hankel transforms are used. More specifically, the correlation is solved as~:
\begin{equation}
	\begin{split}
	&\mc(r,\rho) = \dfrac{1}{\rho^{d-1}}\, \mL^{-1}\left[\mc(r,k)\right]\left[\rho\right], \text{~with~}\\
	& \mc(r,k) := \int_0^{+\infty}\dd \rho \,\rho^{d-1}\,e^{-k\rho}\mc(r,\rho), %= \mL \left[\mc(r,\rho)\right]\left[k\right],
\label{eq:laplace}
	\end{split}
\end{equation}
where  $\mL^{-1}$ denotes the inverse Laplace transform with respect to the pair of variables $\rho,k$. 
From the steady Equations (\ref{eq:steady_wit}-\ref{eq:steady_cit}), the equation on $\mc(r,k^2)$ \footnote{Taking $\mc(r,k^2)$ instead of $ \mc(r,k)$ makes the connection with the calculation of Section 3 particularly apparent.} is now obtained as~:
\begin{equation}
	\begin{split}
	&\left(M_\xi[r] + U_0k^{2} \right) \mc(r,k^2) = \rhs(r,k^2) \\
	&\text{where}\hspace{0.1cm} \rhs(r,k^2) =
%	\begin{cases}
		\dfrac{\phi_\sigma^2}{2^{d-1}\pi} \delta_\epsilon(r) \hspace{0.1cm}\text{( \wit\,case),}\\		
\text{or} \hspace{0.1cm}		&\rhs(r,k^2) = \dfrac{\phi_0}{2^{d-2}\pi} \mc_1(r) e^{-k^2 r/2} \hspace{0.1cm} \text{(\cit\, case).}	
%	\end{cases}.
	\end{split}
	\label{eq:laplacesteady}
\end{equation}

Solving the previous equation and using (\ref{eq:laplace}) to reconstruct the correlation field yields after some routine algebra~:
\begin{widetext}
\begin{equation}
	\begin{split}
	&\mc(r,\rho)=	(1-\xi/2) \,U_0^{-1}\,  \rho^{1-d}\,\left(c_-(r,\rho)+c_+(r,\rho)\right),~~~ \text{with~} \\
%	\text{with~}
 &\mc_-(r,\rho)= \mL^{-1}\left[ K_\omega\left(u^{1/2}\right) \int_0^1 \dd v \,v^{\frac{1+m}{2} -\xi} \rhs\left(rv,\dfrac{u s^2}{\rho}\right) I_\omega\left(u^{1/2}v^{1-\xi/2}\right) \right][s^2], \\
	\text{and~}
 &c_+(r,\rho)= \mL^{-1}\left[ I_\omega\left(u^{1/2}\right) \int_1^{+\infty} \dd v \,v^{\frac{1+m}{2} -\xi} \rhs\left(rv,\dfrac{u s^2}{\rho}\right) K_\omega\left(u^{1/2}v^{1-\xi/2}\right) \right][s^2],
	\end{split}
\label{eq:general_puff}
\end{equation}
\end{widetext}
where $m$ and $\omega$ are the coefficients already referenced in Equation (\ref{eq:coeff}) and the dimensionless parameter $s^2$ is now given by :
\begin{equation}
s^2 = \left(1-\xi/2\right)^2 \dfrac{\rho D_0}{r^2U_0} \left(\dfrac{r}{\lambda}\right)^{\xi}.
\end{equation}
Once again,  $s^2 \sim \tau_{\ball}(\rho)/\tau_\parallel(r)$ is essentially the ratio between the absolute and relative separation Lagrangian time scales, with the former being now the sweeping time-scale, that is  $\tau_{sweep}(\rho) \sim \rho/U_0$.

\subsection{\puff\, \wit\, statistics}
The \puff\, \wit\, correlation field can be computed explicitly, by combining Equations (\ref{eq:laplacesteady}) and (\ref{eq:general_puff}). The final result is :
\begin{equation}
	\begin{split}
		& \mc(r,\rho) = \tilde c_d \left(\dfrac{r}{\epsilon} \right)^{-g} \rho^{1-d}r^{-d} s^{2-2\omega} \exp\left(-\dfrac{1}{4s^2}\right),\\
		&\text{where~~}\tilde c_d = \dfrac{1-\xi/2}{2^{2(\omega+d)} \pi (d+g)}\dfrac{\phi_\sigma^2}{U_0}.
	\end{split}
\end{equation}
Similarly to the Kraichnan case, the control parameter $s^2$ determines the scaling regions. Contrarily to the Kraichnan case, pure scaling is here only present for $s^2 \gg 1$. In that case, the correlation behaves as~:
\begin{equation}
%\mc(r,\rho) \sim  {\phi_\sigma^2 D_0^{1-\omega}}{U_0^{\omega-2}}\left(\dfrac{r}{\epsilon}\right)^{-g } \left(\dfrac{r}{\lambda}\right)^{\xi(1-\omega)} \rho^{2-d -\omega} r^{1-d-\omega},
\mc(r,\rho) \sim  \left(\dfrac{r}{\epsilon}\right)^{-g } \left(\dfrac{r}{\lambda}\right)^{\xi(1-\omega)} \rho^{2-d -\omega} r^{2\omega -2-d},
\end{equation}
%in the is the only parameter that determines the scaling region.
and is obviously very different from  (\ref{wit_asymp}). 
\begin{figure}
\includegraphics[width=0.49\textwidth,trim=0cm 0cm 0cm 0cm,clip]{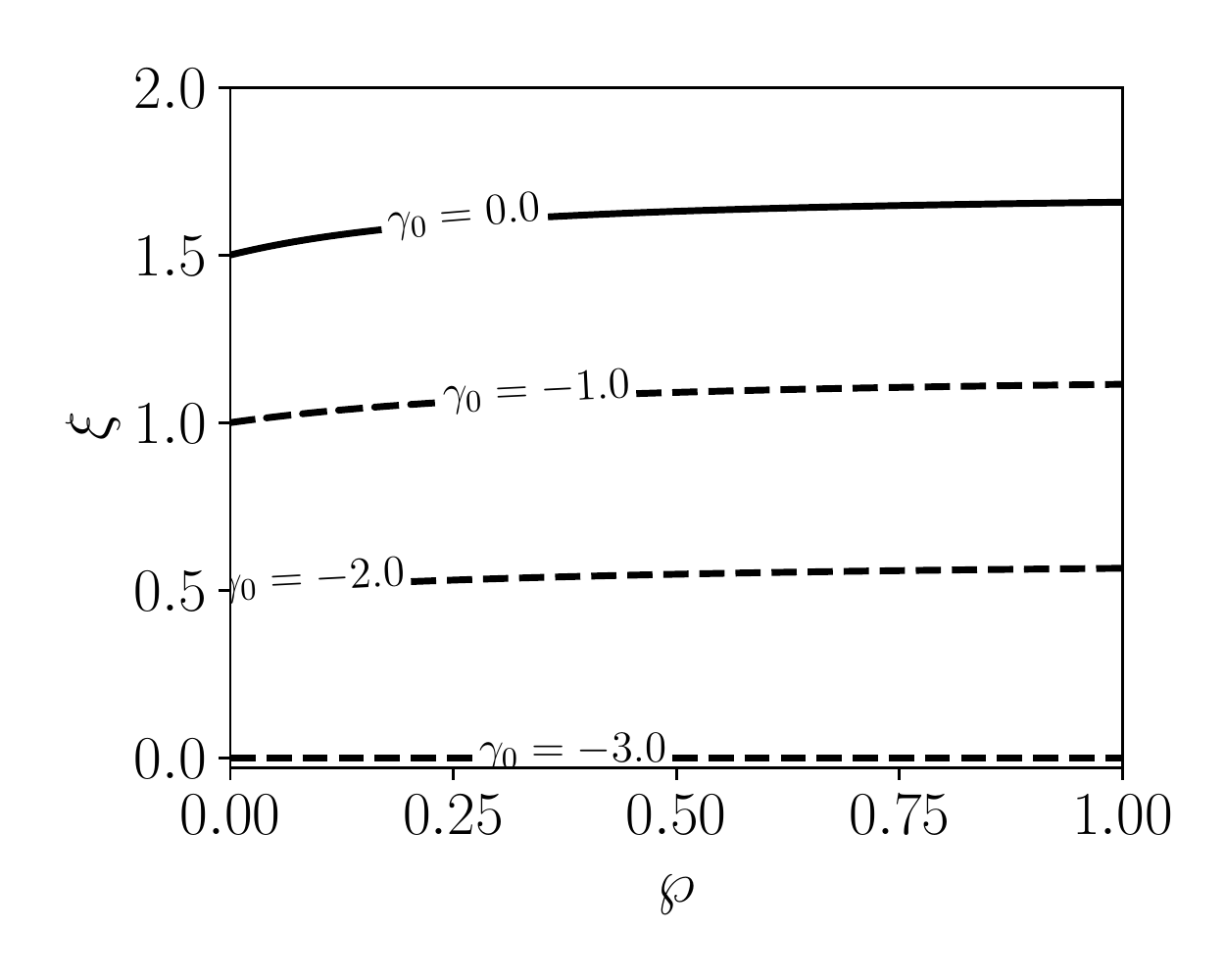}
\includegraphics[width=0.49\textwidth,trim=0cm 0cm 0cm 0cm,clip]{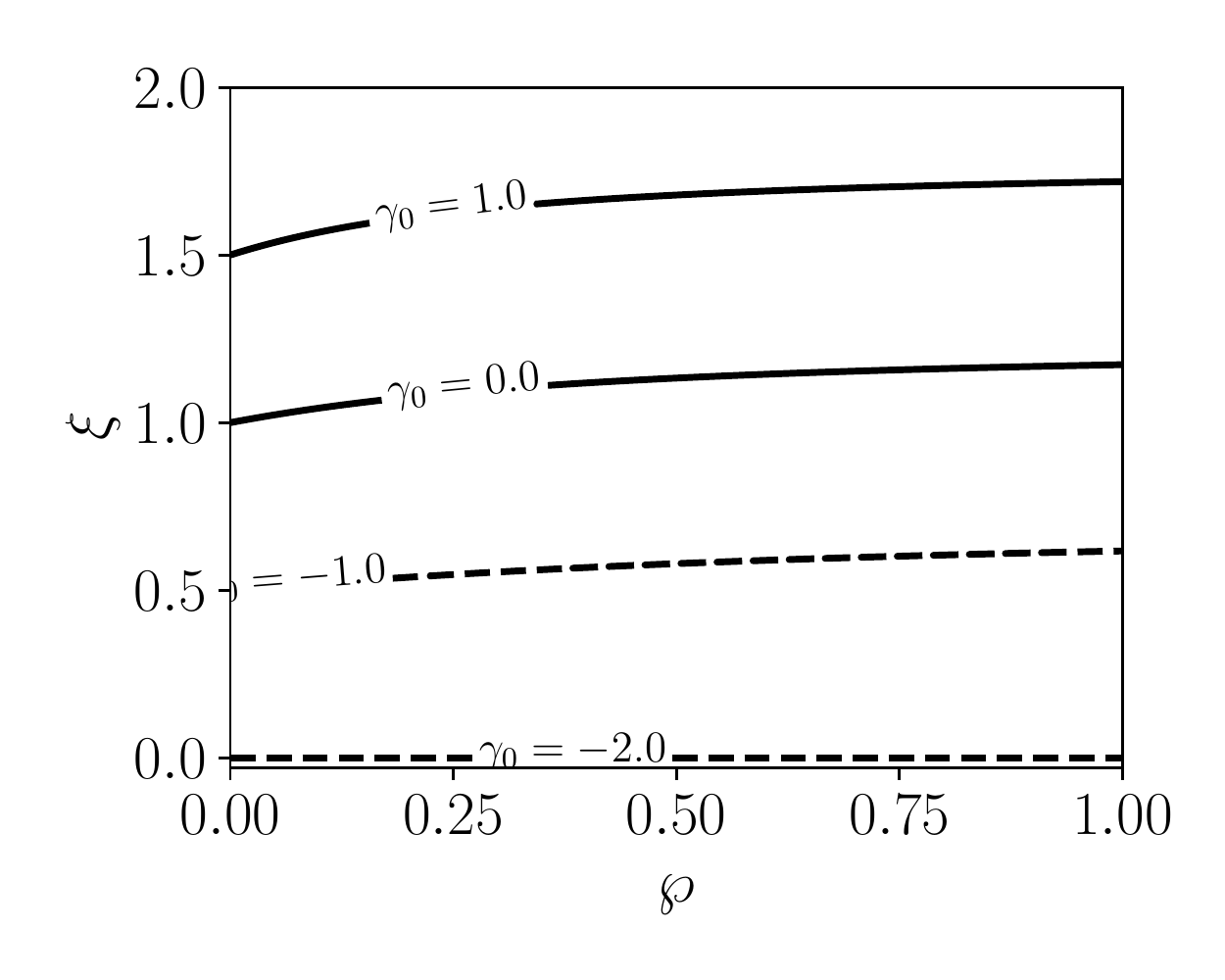}
\caption{The small $r$ scaling exponent $\gamma_0 = d+(f-m-1)/2$ of the quasi-Lagrangian mass in the case of a \cit\, injection, for $d=2$(left) and $d=3 $ (right).}
\label{fig:BEWITcomp}
\end{figure}

The spurious effect of compressibility found in the limit $\epsilon \to 0$ in the Kraichnan case is here still present.
\modif{
The large scale sweeping here translates into the correlation being exponentially damped for large values of $r$. The  competing effects between $\xi$ and $\wp$  that showed up in the Kraichan case for large $r$ is therefore being obliterated.
The small- scale exponent for the quasi-Lagrangian mass is found to be $\gamma_0 = -1-g+(1-\omega)(\xi-2)$, and its behavior is shown on Figure \ref{fig:BEWITcomp}. Not surprisingly, the exponents show little dependence with compressibility. Note that the exponents are negative for small values of $\xi$. However the scaling are \emph{stricto sensu} only valid for $r\gg r_0$, so that $ r_0$ should be taken as a small scale cut-off to make the quasi-Lagrangian mass well defined.}

\subsection{\puff\, \cit\, statistics}
\subsubsection{Average concentration.}
 Combining Equations (\ref{eq:m1ball}) and (\ref{eq:steady_cit}), and prescribing vanishing boundary condition at $\infty$, the average \puff\, concentration is found to be~:
\begin{equation}
\mc_1(|\bx|) = \dfrac{\phi_0}{2^{d-1}\pi U_0 |\bx|^{d-1}}.
\end{equation}
\subsubsection{Correlations.}
The \puff\, correlation is again obtained from  (\ref{eq:general_puff}), observing that $\rhs(rv,us^2/\rho) =  \phi_0 2^{2-d} \pi^{-1}\mc_1(rv) \exp(-uv \Lambda s^2/2)$ with $\Lambda = r/\rho$. The three asymptotic regions previously determined for the Kraichnan case can also be worked out. As for the \puff\, \wit\, case, the statistics in the regions $s \ll1$ are damped by a factor $\exp(-\frac{1}{4s^2})$, and therefore do not display scaling. Only for the case $s \gg 1$, a scaling regime can be identified, namely ~:
\begin{equation}
\begin{split}
&\mc(r,\rho) \sim \dfrac{4}{2^d \omega \pi(2-\xi)} \left(\dfrac{r}{\lambda}\right)^{-\xi} \left(\dfrac{r}{2 \rho}\right)^{\xi-\frac{m \pm f + 3 }{2}}\hspace{-0.4cm}
\mc_1(2 \rho ) \dfrac{\phi_0}{D_0}r^{2}\rho^{-d}, \\
& \text{~~~for  $s\gg1$},
%&~~ \text{for}~~ \pm = \text{sign}({r-2\rho}).
\label{eq:cit_larges_puff}
\end{split}
\end{equation}
where $\pm = \text{sign}({r-2\rho})$.
Up to a constant factor, this expression exactly matches the expression (\ref{eq:cit_larges}) found in the  Kraichnan case, as does the small-scale scaling exponent of the quasi-Lagrangian mass.  The only difference comes from the scaling of the one-point motion, namely $c_1(\rho) \sim \rho^{1-d}$. This is a surprising result, as it suggests that averages over small clouds sizes $r$'s are insensitive to the nature of the absolute/relative interplay.

Let us finally remark that the ballistic behavior of the center of mass destroys the scaling for large $r$'s (region $s \ll 1$), as was already the case in the \wit\, scenario.

\section{Conclusion}
\begin{figure}
\includegraphics[width=0.49\textwidth,trim=0cm 0cm 0cm 0cm,clip]{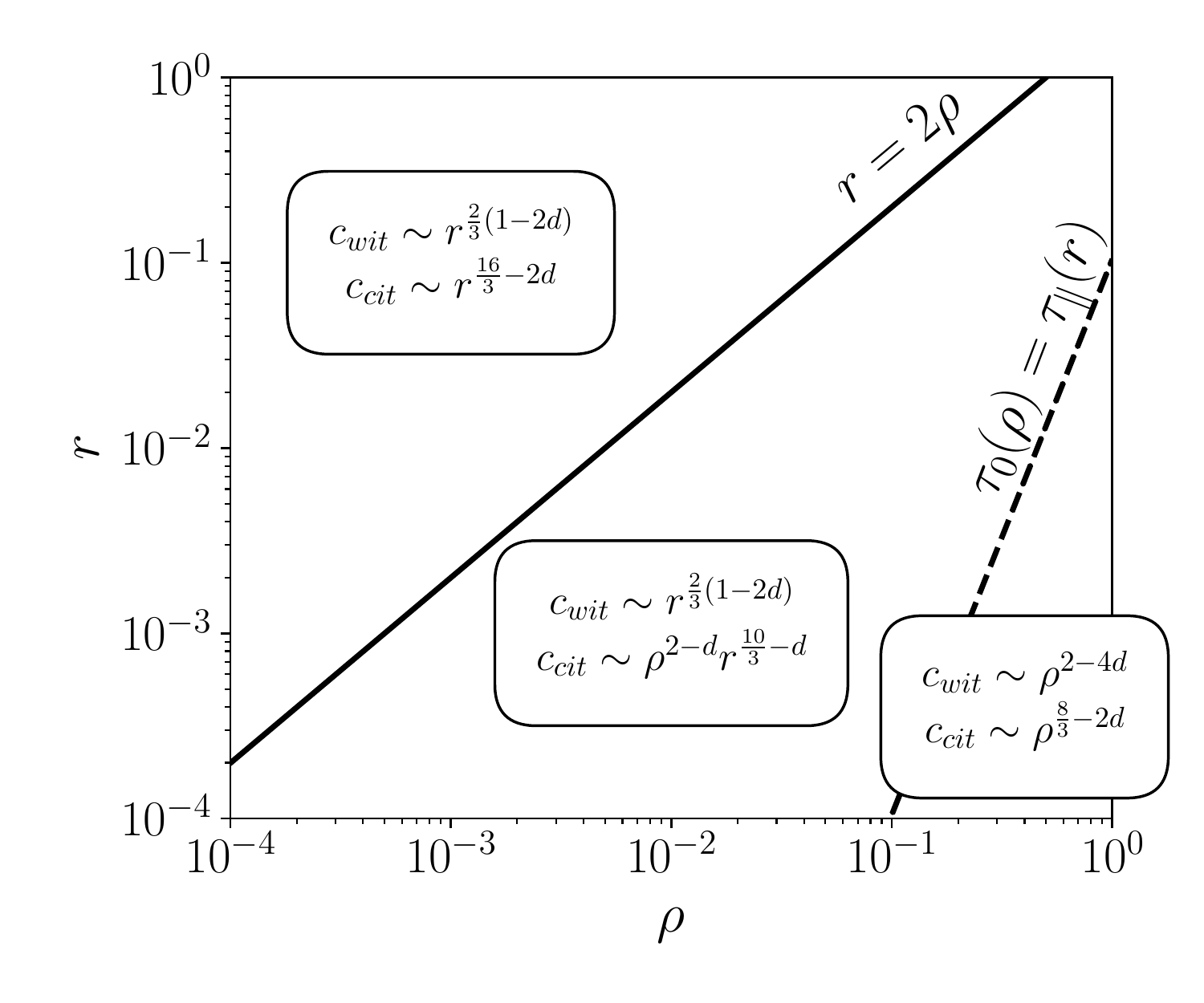}
\includegraphics[width=0.49\textwidth,trim=0cm 0cm 0cm 0cm,clip]{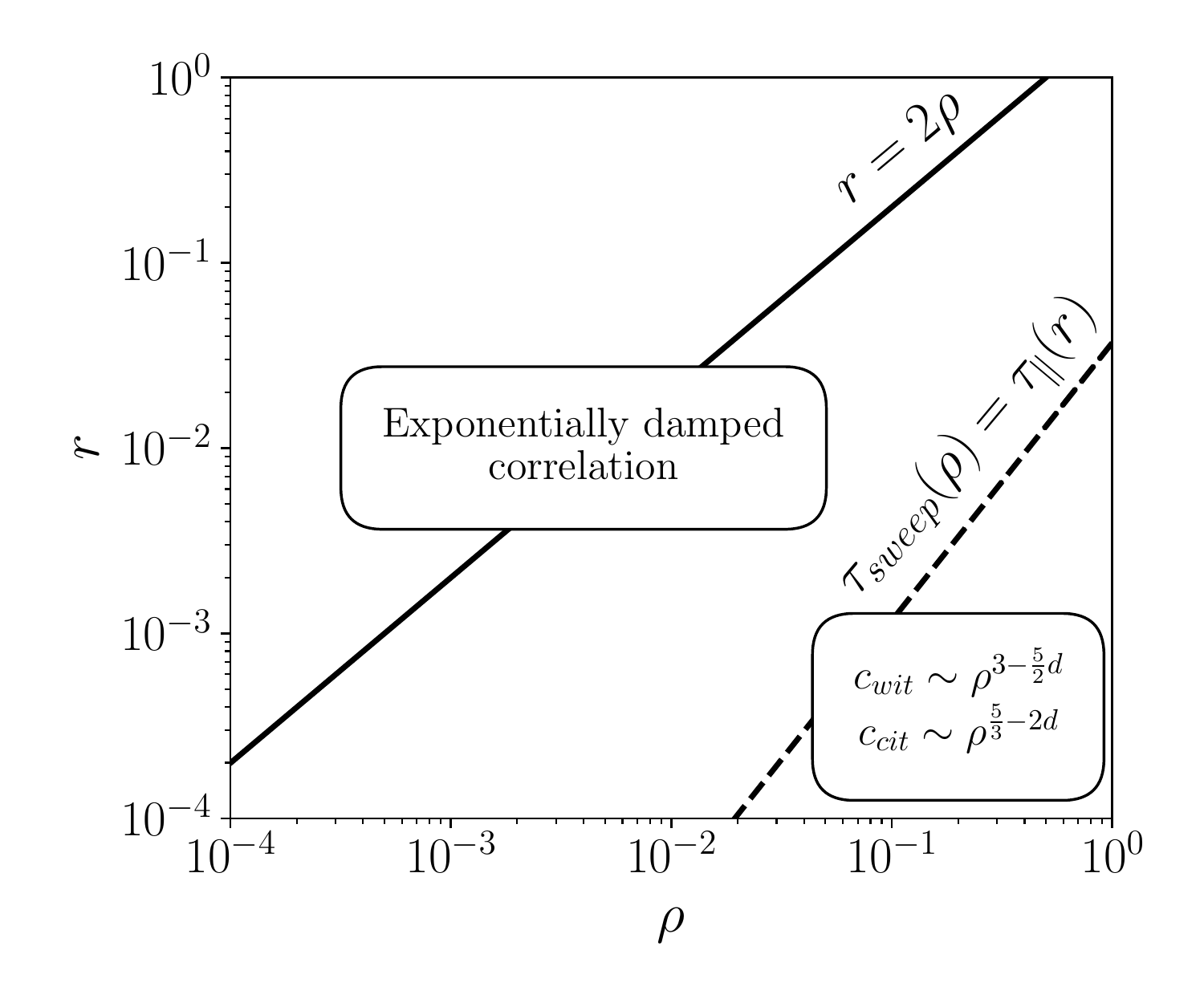}
\caption{Scaling regions of the \wit\, and \cit\, correlation field (here denoted $c_{wit}$ and $c_{cit}$) in the case $\xi=4/3$, relevant to the 2D inverse cascade  and 3D direct cascade, for both the Kraichnan model (top) and the \puff\, model  (bottom). $r$ and $\rho$ are scaled by the large scale $\lambda$. }
\label{fig:sketchscaling}
\end{figure}
\begin{table}
\centering
\def\arraystretch{1.5}
\begin{tabular}{p{2cm}p{5cm}p{5cm}}
%\toprule
&Kraichnan &\puff \\
\hline
\hline
{ $ \bf s^2  \gg 1 $:}&  {$ \rho^{2-d\frac{4-\xi}{2-\xi}}$} & $\rho^{3-d\frac{3-\xi}{2-\xi}} $\\%$ r^{2\xi-4}\rho^{3-d\frac{3-\xi}{2-\xi}} $ \\
{ $ \bf s^2  \ll 1 $:}& {$ r^{2-\xi + \frac{d}{2}(\xi-4)}$}   & No scaling \\
\bottomrule
\end{tabular}
\caption{Scaling of the correlation field for the \wit\, scenario, in the incompressible case $\cp = 0$. Please recall that $s^2  \sim \tau_0(\rho)/\tau_\parallel(r)$ for the Kraichnan model and $s^2  \sim \tau_{sweep}(\rho)/\tau_\parallel(r)$ for the \puff\, model.}
\label{table:witscaling}
\end{table}

\begin{table}
\centering
\def\arraystretch{1.5}
\begin{tabular}{p{2cm}p{5cm}p{5cm}}
%\toprule
$r \ll 2\rho  $&Kraichnan & \puff \\
\hline
\hline
{ $ \bf s^2  \gg 1 $:}& {$  \rho^{4-2d-\xi}$} 
&{$\rho^{3-2d-\xi}$}  \\
{ $ \bf s^2  \ll 1 $:} & {$ r^{2-d+\xi} \rho^{2-d}$} 
& No scaling \\
\bottomrule
\end{tabular}
\begin{tabular}{p{2cm}p{5cm}p{5cm}}
%\toprule
$r \gg 2\rho  $& Kraichnan & \puff \\
\hline
\hline
{ $ \bf s^2  \gg 1 $:} & $r^{2-d-\xi}\rho^{2-d}$ %
 & {$ r^{2-d-\xi} \rho^{1-d}$} \\
{ $ \bf s^2  \ll 1 $:}&  {$r^{4-2d+\xi}$} %
& {No scaling} \\
\bottomrule
\end{tabular}
\caption{Same as Table \ref{table:witscaling}, but  for the \cit\, scenario.}
\label{table:citscaling}
\end{table}

In order to get an overview of the results, some specific scaling behaviors are summarized in Tables \ref{table:witscaling} and  \ref{table:citscaling}, that correspond to the incompressible case ($\cp =0$). Figure \ref{fig:sketchscaling} provides a sketch of the different scaling regions, which are determined by the two parameters $\Lambda = r/\rho$ and $s^2 \sim \tau(\rho) /\tau_\parallel(r)$.\modif{ Please recall that $\tau_\parallel(r) = r^{2-\xi} \lambda^\xi/D_0 $ is essentially the time-scale of relative separation, while $\tau(\rho)$ is the time-scale for the one-point motion, which can be identified to $\tau_0(\rho) = \rho^2/D_0$ (diffusive time-scale) in the Kraichnan ensemble  and $\tau_{sweep}=\rho/U_0$ (sweeping time-scale) in the \puff\, ensemble.}
The salient features are the following :
\begin{itemize}
\item The \cit\, statistics differ from the \wit\, statistics in that they depend on both the Lagrangian timescales ratio $s^2 $ and on the aspect ratio $\Lambda$, while only $s^2$ is relevant for the \wit\, statistics. This observation does not depend on the statistics of the advecting velocity field~:\modif{  It carries through whether the flow is compressible or not, whether $d=2$ or $d=3$, and whether sweeping effects are or not included. This is therefore a robust signature of the injection mechanism itself.}

\item  For small values of $s$, scaling exists. \modif{In both ensemble it is intermittent, in the sense that the inertial scaling of the concentration depends on the large scale $\lambda$.  Note the $\lambda$ dependence is not shown explicitly on Tables \ref{table:witscaling} and \ref{table:citscaling}.}
 The one-point motion affects the correlation when averaged over large shells  ($s \ll1$) in a drastic manner, and is likely to destroy pure scaling behaviors. Physically, this is consistent with the idea that the ballistic motion is faster than the explosive motion for small times. Large clouds correspond to pairs of particles that have essentially spread symmetrically with respect to the source. In the presence of a ballistic one-point motion, the rapid sweeping by the large-scale velocity field make those events extremely unlikely.

\item  Compressibility essentially affects the large $r$ behavior of the correlation, a feature that can be seen from the properties of the quasi-Lagrangian mass. It is here only  explicitly apparent  for the \wit\, statistics in the Kraichnan ensemble. In all the other cases, the effect of compressibility is obliterated by  the continuous injection of mass in the system  and by the large-scale sweeping.

\item In both ensembles, a source with both a \wit\, and a \cit\, contributions induce different scaling behaviors depending on the level of noise in the injection. 
Averages over shells with a small extension $r$  are described by  the limit $s\gg1$, and  the correlation field is there a function of the distance from the source only.  The specific scaling exponents however depend on the type of injection. For a genuine point source, this means that they will depend on levels of fluctuations in the injection.% alter the  scaling exponents. law and  introduce a dependence upon the size $r$. The scaling laws  are non-trivial. We note that the correlation then diverges with vanishing $r$, so that a small amount of noise in the injection is  sufficient to significantly alter the statistics.
\end{itemize}

As emphasized throughout this work, the analytical predictions that are here documented rely heavily on the  white-in-time nature of the underlying prototype turbulent statistics. While this feature is highly unrealistic, it can be hoped that the aforementioned conclusions still hold true  in the presence of non trivial Lagrangian correlations, at least at a qualitative level.%
\modif{%
While the specific values of the scaling regimes that were here found in this work are hardly likely to be seen in a real flow, it could be expected that some robust features might carry trough. For instance in the incompressible, Figure \ref{fig:sketchscaling} shows that for a velocity field with Kolmogorov-like scaling, one should be able to distinguish between regions where the statistics of the correlations are dominated by one point motion, and where no dependence on r is shown ($s\gg1$) from regions dominated by relative separation where on the contrary no dependence on $\rho$ is shown ( $s \ll $1).
} 

In  finite-Reynolds-number turbulent flows, both relative and absolute dispersion however have multiple stages. Whether scaling regimes for the concentration field are indeed to be found  is not granted.
 This is an open question that would benefit from being investigated using either Direct Numerical Simulations or laboratory experiment in the light of the present framework.   
For realistic injection mechanisms,  it is to be tested whether scaling regimes of the fluctuation field depend on the level of noise in the injection mechanism, as implied for instance by  Figure \ref{fig:sketchscaling}. It is  to be seen whether the statistics  have connections with either the \puff\, or the Kraichnan statistics.

\modif{In practice, one might also wish to consider positive-definite random sources, that emits puffs of particles randomly in time, but unlike our \wit\, injections  do not remove any. This situation can in fact be checked to be  ``intermediate'' between the \wit\, and the \cit\, case~:  The average concentration $\mc_1$ is non-zero and prescribed by the average \cit\, concentration, while the correlation field is prescribed by the  \wit\, correlation field.}

The approach described in this paper could also naturally be extended to study non ideal turbulent transport, either involving  inertial or active particles as initiated by \cite{afonso_point-source_2011}, or involving  non isotropic turbulent statistics.

\begin{acknowledgments}
 I acknowledge insightful discussions  with  J\'er\'emie Bec,  Giorgio Krstulovic and Fran\c cois Laenen.
I also thank ICTS-TIFR for their hospitality that led to final completion of this work~: In particular the ICTS programs ICTS/taly2018/01 and  ICTS/ispcm2018/02 as well as the support from the  DST (India) project ECR/2015/00036.
%
%Major revisions to this work were added while visiting ICTS Bangalore, with support from the Indo-French Centre for Applied Mathematics .
\end{acknowledgments}
%\newpage	

\appendix
\section{ Computation of the correlation field}
\label{sec:details}
This appendix contains some  details about the algebra involved in the computation of the correlation fields, namely \emph{(1)} the derivation  of the mode-to-mode equation (\ref{eq:hankelsteady}),  \emph{(2)} the general solution (\ref{eq:general}), \emph{(3)} the asymptotics of the \cit\, correlation.

\subsection{Hankel transforms of the steady states equation.}
\label{sec:HankelDetails}
Equation (\ref{eq:hankelsteady}) is obtained from the steady states equations (\ref{eq:steady_wit}-\ref{eq:steady_cit})~: 
\begin{equation*}
	\begin{split}
	&\mM_2[\bx,\bx^\prime] \mc(r,\rho) = \rhs(\bx,\bx^\prime) \text{~~with}\\
& \rhs(\bx,\bx^\prime) =
\begin{cases}
& \phi_\sigma ^2 \delta(\bx)\delta(\bx^\prime) \hspace{0.5cm}\text{(WIT)}\\
& \phi_0 \left(\mc_1(\bx^\prime) \delta(\bx) + \mc_1(\bx) \delta(\bx^\prime)\right) \hspace{0.5cm}\text{(CIT)} \\
\end{cases},
	\end{split}
\end{equation*}
where we recall that $r(\bx,\bx^\prime) := |\bx - \bx^\prime|$ and $\rho(\bx,\bx^\prime) := |\bx + \bx^\prime|/2$, and that $ \mM_2 = \mM_1[\brho] + \mM_2[\br]$ from Equation (\ref{eq:M2_asymp}).

Hankel transforming both sides of the previous equation yields :
\begin{itemize}
\item for the left-hand side :
\begin{align*}
	\text{lhs}(r,k) &= \left(\dfrac{2}{\pi}\right)^{d-2}\, \hspace{-0.3cm}\int_0^{\infty}\dd \rho \,\rho^{d-1} \mJ_0(k\rho)\,\left(\mM_1[\rho]+\mM_\xi(\br) \right) \mc(r,\rho) \\
&= \left(\mM_\xi(\br) + \dfrac{D_0}{2}k^2\right) \mc(r,k),
\end{align*}
where the second line comes from a double integration by parts with respect to $\rho$, and from the $\mJ_0$'s being the isotropic eigen-functions of the diffusion operator $\mM_1$;
\item for the right-hand side :
\begin{align*}
	\rhs(r,k) &= \left(\dfrac{2}{\pi}\right)^{d-2}\,\int_0^{\infty}\dd \rho \,\rho^{d-1} \mJ_0(k\rho) \rhs(\bx,\bx^\prime)\\
	 &= \lim_{\epsilon \to 0} \left(\dfrac{2}{\pi}\right)^{d-2}\, \hspace{-0.4cm}  \dfrac{1}{2^{d-1}\pi}\int_0^{\infty}\dd \bx \dd \bx^\prime \delta_{\epsilon}\left(r-r(\bx,\bx^\prime)\right) \,\rho^{d-1} \\
&\hspace{5cm}\mJ_0(k\rho(\bx,\bx^\prime)) \rhs(\bx,\bx^\prime),
\end{align*}
where the notation $\delta_{\epsilon}(r)$ essentially denotes a compact-support approximation to a radial Dirac distribution, namely a step function  that  vanishes for $r>\epsilon$ and otherwise takes the constant value $\epsilon^{-d} V_d^{-1}$, where $V_d$ is the volume of the unit sphere in dimension $d$.
\end{itemize}
Equation (\ref{eq:hankelsteady}) follows. For the \wit\, statistics, the final result involves a Dirac distribution for the right-hand-side. To avoid any confusion, we find it safer to keep track of the ``source extension'' $ \epsilon \ll 1$, and not take directly the limit ``$\epsilon \to 0$'', hence the $\epsilon$ subscript in Equation (\ref{eq:hankelsteady}).

\subsection{General form of the fluctuation field}
The general solution (\ref{eq:general}) is obtained by solving Equation (\ref{eq:hankelsteady}) explicitly and transforming $\mc(r,k)$ back into $\mc(r,\rho)$.
To solve for $\mc(r,k)$, one first works out the isotropic contribution to the operator $\mM_\xi[\br]$ as 
\begin{equation*}
	\begin{split}
	&\mM_\xi[\br] = -D_0 \left(\dfrac{r}{\lambda}\right)^\xi\left(\partial^2_{rr} +\dfrac{m(\cp,\xi)}{r}\partial_{r} + \dfrac{n(\cp,\xi)}{r^2}  \right) , \text{~~where}\\
	& m(\cp,\xi) = (d+\xi-1)\left( 1 + \dfrac{\cp \xi}{1+\cp \xi}\right) ~~\text{and}\\
	&n(\cp,\xi) = (d+\xi-2)\left(d +\xi \right) \dfrac{\cp \xi}{1+\cp \xi}.
	\end{split}
\end{equation*}
One may observe that a pair of independent homogeneous solutions of (\ref{eq:general}) is  
\begin{equation*}
	\begin{split}
	&\phi_I(r,k) = r^{(1-m)/2}I_\omega\left(\eta k r^{1-\xi/2}\right),\\
	& \text{and}~ \phi_K(r,k) = r^{(1-m)/2}K_\omega\left(\eta k r^{1-\xi/2}\right),\\
	&\text{where} ~\omega = \dfrac{\left((m-1)^2-n\right)^{1/2}}{2-\xi} \text{~and~} \eta = \dfrac{\sqrt 2}{2-\xi}\lambda^{\xi/2},
	\end{split}
\end{equation*}
and where $I_\omega$ and $K_\omega$ are the modified Bessel of the first and second kind.
The  solution $\mc(r,k)$ is then obtained by a brute-force use of the ``variation of the constant'' method, which yields
\begin{equation*}
\begin{split}
\mc(r,k) = \dfrac{\lambda^\xi}{D_0(1-\xi/2)} &\left(  \phi_I(r,k)  \int_r^\infty \dd r^\prime\, {r^\prime}^{m-\xi} \phi_K(r^\prime,k) \rhs(r^\prime,k)  \right. \\
&\hspace{0.1cm}\left. + \,\, \phi_K(r,k) \int_0^r \dd r^\prime\, {r^\prime}^{m-\xi} \phi_I(r^\prime,k) \rhs(r^\prime,k) \right).
\end{split}
\end{equation*}

Reconstructing the fluctuation field as $c(r,\rho) = \int_0^\infty \dd k\, k^{d-1}  c(r,k) \mJ_0(k\rho)$, and performing the change of variables 
``$v=r^\prime/r$ and $u = \eta  r^{1-\xi/2} k$'' yields the looked-for general expression (\ref{eq:general}).

\subsection{\cit\, asymptotics}
\label{eq:citdetails}
The \cit\, asymptotics (\ref{eq:cit_larges})-(\ref{eq:cit_smallssmallL}) are obtained from (\ref{eq:general}) by \emph{(i)} approximating the modified Bessel functions $I_\omega, K_\omega$ with their asymptotic behaviour, and \emph{(ii)}  integrating  over the $u$- before the $v$-variable. After changing $u \to 2 u/(\Lambda s)$, one gets from Equation (\ref{eq:general}), in the limit  $s\gg 1$~:
\begin{widetext}
\begin{align*}
	c_-(r,\rho) &\simeq \dfrac{\phi_0}{2 \pi^{d-1} \omega s^d} \left(\dfrac{2}{\Lambda}\right)^{d}   \int_0^1 \dd v v^{\frac{1+m+f}{2} -\xi } \mc_1(rv) \, \int_0^{+\infty} \dd u\, u^{d-1} \mJ_0\left(u \dfrac{2}{\Lambda}\right)\mJ_0\left(u v\right),\\
	\text{and}~c_+(r,\rho) &\simeq \dfrac{\phi_0}{2 \pi^{d-1} \omega s^d} \left(\dfrac{2}{\Lambda}\right)^{d}   \int_1^\infty \dd v v^{\frac{1+m-f}{2} -\xi } \mc_1(rv) \, \int_0^{+\infty} \dd u\, u^{d-1} \mJ_0\left(u \dfrac{2}{\Lambda}\right)\mJ_0\left(u v\right).
\end{align*}
\end{widetext}

The result (\ref{eq:cit_larges}) is then obtained by observing that $$\int_0^\infty \dd u u^{d-1} \mJ_0(ux)\mJ_0(uy) = (\pi/2)^{d-2} (xy)^{1-d/2}\delta(x-y), $$ with $\delta$ here denoting the one-dimensional Dirac distribution.\\

Similarly, in the limit  $s\ll 1$~:
%\begin{widetext}
\begin{align*}
	c_-(r,\rho) &\simeq \dfrac{\phi_0 s^{1-d}}{2 \pi^{d-2}}\int_0^1 \dd v v^{m/2 -3 \xi/4 } \mc_1(rv) \, \int_0^{+\infty} \dd u\, u^{d-2} \mJ_0\left(u \right) \mJ_0\left(u v \dfrac{\Lambda}{2} \right) e^{-\frac{u}{s}\left(1-v^{1-\xi/2} \right)},\\
	c_+(r,\rho) &\simeq \dfrac{\phi_0 s^{1-d}}{2 \pi^{d-2}}\int_1^\infty \dd v v^{m/2 -3 \xi/4 } \mc_1(rv) \, \int_0^{+\infty} \dd u\, u^{d-2} \mJ_0\left(u \right) \mJ_0\left(u v \dfrac{\Lambda}{2} \right) e^{-\frac{u}{s}\left(v^{1-\xi/2}-1 \right)}.
\end{align*}
%\end{widetext}

Explicit expressions for $\mc_\pm$ can be obtained in the limits $\Lambda  \gg 1$ and $\Lambda \ll 1$. Noticing that
$\int_0^\infty \dd u \, \mJ_0(uv) \exp(-|x| u) = \left(x^2+v^2 \right)^{(1-d)/2}$, we obtain the asymptotic behaviour 	for $\Lambda \ll 1$~:
\begin{equation*}
	\begin{split}
	c_-+c_+ & \simeq \dfrac{\phi_0}{2 \pi^{d-2}}s^{1-d} \int_0^{+\infty} \dd v \dfrac{v^{m/2-3 \xi/4} }{\left(1+\dfrac{X(v)^2}{s^2}\right)^{(d-1)/2}}\\
%	& \hspace{0.1cm} \text{for $\Lambda \ll 1$}.\\
	\end{split}
\end{equation*}
Similarly for $\Lambda \gg 1$, the asymptotic behaviour is~:
\begin{equation*}
	\begin{split}
	c_-+c_+ & \simeq \dfrac{\phi_0}{2 \pi^{d-2}}\left(s \dfrac{2}{\Lambda}\right)^{1-d} \int_0^{+\infty} \dd v \dfrac{v^{m/2-3 \xi/4} }{\left(v^2+\dfrac{4}{\Lambda^2} \dfrac{X(v)^2}{s^2}\right)^{(d-1)/2}} \\
%& \hspace{0.1cm} \text{for $\Lambda \gg 1$},
	\end{split}
\end{equation*}
where $X(v) := |1-v^{1-\xi/2}|$.\\
Since $s \ll 1 $, the behaviours of those  integrals are dominated by the behaviours near $v=1$, for instance :
\begin{align*}
\int_0^{+\infty} \dd v  \dfrac{\mc_1(rv) \,v^{m/2-3 \xi/4} }{\left(1+\dfrac{X(v)^2}{s^2}\right)^{(d-1)/2}} & \sim 2 \mc_1(r)\int_{1-\epsilon}^1  \, \dfrac{\dd v}{\left(1+ \dfrac{X(v)^2}{s^2}\right)^{(d-1)/2}}, \\
&  \hspace{0cm} \sim %
2 \mc_1(r) \dfrac{s}{1-\xi/2} \chi_d(s),
%\begin{cases}
%&2 \mc_1(r) \dfrac{s}{1-\xi/2} \text{atan} \dfrac{1-\xi/2}{s}\epsilon \hspace{0.4cm}\text{for $d=3$}\\
%&  -2 \mc_1(r) \dfrac{s}{1-\xi/2} \log s\hspace{0.4cm}\text{for $d=2$}.
%\end{cases}
\end{align*}
where $\chi_d(s) = \text{asinh}( (1-\xi/2)\epsilon/s) \sim -\log s $ for $d= 2$ and $\chi_d(s) = \text{atan}( (1-\xi/2)\epsilon/s) \sim \Pi/2 $ for $d= 3$.  
The behaviours (\ref{eq:cit_smallssmallL}) and  (\ref{eq:cit_smallslargeL}) follow.
The asymptotic behaviour (\ref{eq:cit_larges_puff}) valid for the \puff\, ensemble is obtained along the same line.
\bibliographystyle{apsrev4-1}
\bibliography{PointSource_FULL}
\end{document}